\documentclass[journal,comsoc,twopage,final]{IEEEtran}
\usepackage[utf8]{inputenc}
\usepackage[T1]{fontenc}
\pdfoutput=1
\usepackage[usenames,x11names,svgnames]{xcolor}
\usepackage[hyphens]{url}
\usepackage[breaklinks,hidelinks]{hyperref}
\usepackage{graphicx}
\usepackage{amsmath}
\usepackage[cmintegrals]{newtxmath}
\usepackage{cite}
\usepackage{array}
\usepackage{amssymb}
\usepackage{xspace}
\usepackage{listings}
\usepackage{mdwlist}
\usepackage{booktabs}
\ifCLASSOPTIONcompsoc
  \usepackage[caption=false,font=normalsize,labelfont=sf,textfont=sf]{subfig}
\else
  \usepackage[caption=false,font=footnotesize]{subfig}
\fi
\usepackage{stfloats}
\usepackage[export]{adjustbox}
\usepackage{csquotes}

\newcommand{\fref}[1]{Fig.~\ref{#1}}
\newcommand{\sref}[1]{Section~\ref{#1}}

\newcommand{\tref}[1]{Table~\ref{#1}}

\newcommand{\socketintent}{{Socket~Intent}\xspace}
\newcommand{\socketintents}{{Socket~Intents}\xspace}
\newcommand{\socketintentpolicy}{{Socket~Intents~Policy}\xspace}
\newcommand{\socketintentspolicy}{{Socket~Intents~Policy}\xspace}
\newcommand{\socketintentpolicies}{{Socket~Intents~Policies}\xspace}
\newcommand{\polmgr}{{Multi~Access~Manager}\xspace}
\newcommand{\prototype}{{Socket~Intents Prototype}\xspace}

\newcommand{\BSDsockets}{BSD Sockets\xspace}
\newcommand{\vBSDsockets}{vanilla BSD Sockets\xspace}
\newcommand{\VBSDsockets}{Vanilla BSD Sockets\xspace}

\newcommand{\BSDsocketAPI}{BSD Socket API\xspace}
\newcommand{\vBSDsocketAPI}{vanilla BSD Socket API\xspace}
\newcommand{\BSDsocketinterf}{BSD Socket Interface\xspace}

\newcommand{\filesizeIntent}{\textsl{Size to be Received Intent}\xspace}

\newcommand{\poln}[1]{{\fontfamily{pzc}\selectfont#1}\xspace}

\newcommand{\eafpolicy}{\poln{EAF} policy\xspace}

\newcommand{\fcall}[1]{\texttt{#1{\small()}}\xspace}
\newcommand{\lname}[1]{\texttt{#1}\xspace}

\let\underscore\_
\newcommand{\myunderscore}{\renewcommand{\_}{\underscore\hspace{0pt}}}
\myunderscore



\hypersetup{
        pdftitle={Socket Intents: OS Support for Using Multiple Access Networks and its Benefits for Web Browsing},
        pdfauthor={{Philipp S. Tiesel <philipp@inet.tu-berlin.de>}, {Theresa Enghardt <theresa@inet.tu-berlin.de>}, {Mirko Palmer <mirko@inet.tu-berlin.de}, {Anja Feldmann <anja@inet.tu-berlin.de>}},
        pdfkeywords={Access Selection, Socket Intents, BSD Sockets, Application Awareness, Web Performance}
}

\begin{document}

\title{Socket Intents: OS Support for Using Multiple Access Networks and its Benefits for Web Browsing}

\author{Philipp S. Tiesel, Theresa Enghardt, Mirko Palmer, Anja Feldmann}

\markboth{Tiesel \MakeLowercase{\textit{et al.}}: Socket Intents: OS Support for Using Multiple Access Networks and its Benefits for Web Browsing}{}

\maketitle

\thispagestyle{empty}

\begin{abstract}
In today's Internet, mobile devices are connected to multiple access networks, e.g., WiFi/DSL and LTE.
To take advantage of the networks' diverse paths characteristics (delay, bandwidth, and reliability) and aggregate bandwidth, we need smart strategies for choosing which interface(s) to use for what traffic.
In this paper, we present an approach how to tackle this challenge as part of the Operating System (OS): 
With the concept of \socketintents, applications can express what they \emph{know} about their communication pattern and their preferences.
Using our \prototype and our modified \BSDsocketinterf, this information is used to  choose the most appropriate path or path combination on a per message or per connection basis.
We evaluate our system based on the use case of Web browsing:
Using our prototype and a client-side proxy, we show the feasibility and benefits of our design.
Using a flow-based simulator and a full factorial experimental design, we study a broad range of access network combinations (based on typical DSL and LTE scenarios) and real workloads (Alexa Top~100 and Top~1000 Web Sites).
Our policies achieve performance benefits in more than 50\% of the cases and speedups of more than factor two in 20\% of the cases without adding overhead in the other cases.

\end{abstract}


\section{Introduction}
\label{sec:intro}

\IEEEPARstart{T}{oday,} mobile devices can usually access the Internet over more than one access network.
For example, mobile phones often have the choice between WiFi and cellular networks.
By using the paths provided by these networks at the same time, it is possible
to aggregate their bandwidth or switch between them to increase overall network availability.
Moreover, applications can take advantage of the different characteristics of
these paths, e.g., delay, bandwidth, and expected availability, by using the
most suitable path(s) according to the application's demands.

To illustrate why these characteristics matter, we take a look at two use-cases:
If a user fetches news headlines or stock market quotes, quickly loading small
objects ensures responsiveness; so the user prefers a path with low latency.
If a user streams a TV series over HTTP, high throughput is most important
to provide high video quality.
Therefore, if the available paths vary in bandwidth and latency, both use-cases benefit from assigning their communications to the path with the most suitable characteristics.
Similarly, there is often a choice between multiple destinations, e.g., CDN nodes hosting the same content. 

Despite the possible benefits, the available diversity is usually not exploited. 
While the application knows its demands, selecting the path and destination within the application is impractical most of the time. 
This stems from the fact that information required for an appropriate selection
is often not available, such as detailed characteristics of the available paths or information about cross-traffic from other applications.
In addition, implementing path selection strategies is a rather complex task which touches resource management and, thus, belongs to the domain of the operating system (OS).
At the OS level, all necessary information including path characteristics and traffic information are available. 
But when a socket gets connected, the OS is not able to distinguish between different application demands.
Consequently, there is no way for the OS to know what to optimize for. 

So far, application-aware OS support for multiple paths, the focus of this paper, is still very limited.
Most related research~\cite{aijaz2013survey,balasubramanian2010augmenting,vallina2012david}
or commercial systems~\cite{muacc-csws12,Rebecchi-Offloading} focus on WiFi 
off- or onloading at the network layer or application layer for well-defined use cases.
Multi-Path aware transport protocols, including SCTP~\cite{rfc4960} and Multipath TCP (MPTCP)~\cite{Paasch:2014:MT:2578508.2591369,rfc6182} allow to aggregate bandwidth across multiple access networks and provide fallback in case of network failures, but are agnostic to applications' needs.

To this end, we introduced the concept of \emph{Socket Intents}~\cite{socketintents}.
Socket Intents allow applications to share their knowledge about their communication pattern and express performance preferences in a generic and portable way.
Thus, an application developer can inform the OS about what the intent of the communication is and what she knows about the communication:
\begin{itemize}
    \item \textbf{Preferences} whether to optimize for bandwidth, latency, or cost 
    \item \textbf{Characteristics} expected packet rates, bitrates or size of the content to be sent or received.
    \item \textbf{Expectations} towards path availability or packet loss
    \item \textbf{Resiliences} whether the application can gracefully handle certain error cases
\end{itemize}
None of these are hard requirements, e.g., transport protocol guarantees or QoS style reservation. However, they are crucial for the OS to do path and destination selection on behalf of the application.

In this paper, we demonstrate \emph{how Socket Intents can be implemented} on top of \vBSDsockets by designing a system that enables automated path and destination selection within the OS and evaluate the \emph{impact of Socket Intents on Web performance}.

Our prototype implementation extends the \BSDsocketAPI to support path and destination selection for two communication granularities: streams and messages.
It demonstrates the feasibility of automated path and destination selection within the OS but also reveals major limitations of the \BSDsocketAPI.

\pagebreak
The main contributions of this paper are:
\begin{itemize*}
  \item We extend our original prototype~\cite{socketintents} to support message granularity communication units, e.g., HTTP requests.
    Guided by the application needs provided by \socketintents and path characteristics, we realize connection caching and implicit connection pools in the OS.
    In addition, we enable our prototype to take control over MPTCP.

  \item We introduce the \emph{Earliest Arrival Time} (\poln{EAF}) policy as informed path selection strategy for Web objects.
    Based on the information provided by the \filesizeIntent, \poln{EAF} assigns each request to the path it is predicted to complete on first.

  \item We demonstrate applicability of our prototype by implementing a client side HTTP proxy. With only 20 lines of additional code, the proxy takes advantage of connection caching, \socketintents and the \eafpolicy .

  \item We evaluate the benefits of using \socketintents with the \eafpolicy in two ways: 
    A small testbed study using our proxy and an extensive simulator study using a custom flow-based simulator.
    Our simulation uses a full factorial experimental design and covers the Alexa Top~100 and Top~1000 Web sites over a wide range of network characteristics resembling typical residential broadband and cellular network characteristics. 
\end{itemize*}

\section{Socket Intents Concept}
\label{sec:intents}

To perform path and destination selection within the OS, the OS needs to know
what to optimize for -- the application demands.
Therefore, we introduced the concept of \emph{Socket Intents}~\cite{socketintents}.
Socket Intents allow applications to share their knowledge about their communication patterns and express performance preferences in a generic and portable way.
Intents are hints for the OS, pieces of information, that allow an application programmer to express what they know about the application's needs or intentions for each communication unit.
They indicate what the application wants to achieve, knows, or assumes.
In contrast to transport features or QoS-style reservations, they are not requirements but only considered in a \emph{best-effort} manner, e.g., as input to path and destination selection heuristics within the OS. 
Possible intents, as shown in \tref{tab:intents}, include \emph{Traffic Category}, \emph{Size to be Sent/Received}, \emph{Timeliness}, \emph{Duration}, or \emph{Resilience of connectivity}. 

Applications have an incentive to specify their intents as accurately as possible to take advantage of the most suitable resources. 
We expect applications to selfishly specify their preferences.
Since the OS knows about the available network resources and the intents of multiple applications, it can balance the different requirements and penalize misbehaving applications.

Socket Intents are independent of the actual Socket API and can be applied to message granularity communications, e.g., UDP messages or HTTP requests, as well as stream granularity communications, e.g., TCP connections.
The information provided by the application is structured as key-value-pairs.
The key is a simple string representing the type of a Socket Intent. Values can be represented as an \emph{enum}, \emph{int}, \emph{float}, \emph{string}, or a {sequence} of the aforementioned data types.
\tref{tab:intents} gives an overview of Socket Intent types as specified in our recent IETF draft~\cite{I-D.tiesel-taps-socketintents}.
Despite the variety of Intents we define in this section, the remainder of this paper focuses on how to realize Socket Intents as an extension to the \BSDsocketAPI and the benefits of using the \filesizeIntent.

\begin{table}[t!]
    \caption{Socket Intents Types}
    \label{tab:intents}
    \centering
\resizebox{0.95\columnwidth}{!}{%
\small
\begin{tabular}{llcccc}
    \toprule
    \textbf{Intent Type}        & \textbf{Data Type}    & \multicolumn{2}{l}{\textbf{Applicable Granularity}}                               \\
                                &                       & \textbf{Message}  & \textbf{Stream}  \\
    \midrule                                                                                   
    Traffic Category            & Enum                  &                   & \checkmark       \\
    Size to be Sent             & Int (bytes)           & \checkmark        & \checkmark       \\
    Size to be Received         & Int (bytes)           & \checkmark        & \checkmark       \\
                                                                                            
    Duration                    & Int (msec)            &                   & \checkmark       \\
    Bitrate Sent                & Int (bytes/sec)       &                   & \checkmark       \\
    Bitrate Received            & Int (bytes/sec)       &                   & \checkmark       \\

    Burstiness                  & Enum                  &                   & \checkmark       \\
    Timeliness                  & Enum                  & \checkmark        & \checkmark       \\
    Disruption Resilience       & Enum                  & \checkmark        & \checkmark       \\
    Cost Preferences            & Enum                  & \checkmark        & \checkmark       \\
    \bottomrule
\end{tabular}%
}

\end{table}


\section{Challenges Imposed by \BSDsockets} 
\label{sec:bsdsockets_challenges}

In Unix-like OSes, \BSDsockets are the standard interface between applications and the network stack.
Typically, applications that want to connect to a server first resolve the server's hostname using \fcall{getaddrinfo}, then create a socket file descriptor using \fcall{socket} and call \fcall{connect} to establish the connection.
To each of these calls information obtained from \fcall{getaddrinfo} is passed.

With \VBSDsockets, taking advantage of multiple paths or choosing among several
destinations is complicated.
One reason is that the \BSDsocketAPI designers considered multi-homed hosts a corner case.
The \fcall{bind} socket call allows applications to choose the source address of an outgoing communication\footnote{
Otherwise, the OS uses the IP address of the interface via which it routes to the given destination.}.
If the system is configured with a routing policy to send traffic with a specific source address over an associated paths, application can set the source address to implicitly choose the outgoing interface and next-hop and, therefore, large portions of the path.

Besides this hack, \VBSDsockets do not offer support for multiple access networks:
Applications that want to use multiple interfaces usually have to have their own heuristics for selecting paths.
Choosing among paths is difficult as the necessary information is often
difficult to gather and may require special privileges. Moreover, it differs greatly by Unix flavor.

\label{sub:dnsifbind}
Another complication occurs when selection a destination:
When resolving the hostname to obtain a destination address, applications need
to ensure not to mix results for the same hostname resolved via a different interface.
For example, CDNs and major Web sites often rely on DNS-based server selection and load balancing.
These mechanisms are most useful if the DNS query is sent via the same interface as the actual traffic.
If the application sends the traffic over another interface, the chosen server may be suboptimal, which can lead to significant performance degradations.
Yet, the resolver library of the \vBSDsocketAPI does not allow us to isolate results acquired via multiple paths.
For a more detailed discussion, see \cite{ANRW-multihomed-singlelink}.

Furthermore, the communication units used by \vBSDsocketAPI are implied by the transport protocol and must match the socket type passed to the \fcall{socket} call.
Thus, for stream-based communication protocols like TCP, the application can only choose a path and endpoint for the whole stream.
But communication units of actual applications are often not aligned with the
communication granularity of the transport protocol.
For example, requests in HTTP ---the dominant protocol on the
Internet~\cite{httpNarrowWaist,prichterIXP}--- correspond to a message based communication performed over a stream transport.
An HTTP based application can choose for each request to either open a new TCP connection or reuse an existing one.
The former allows choosing among multiple interfaces using \fcall{bind}.
The latter saves 2\,RTTs for the TCP handshake, a few 100~KB for the TLS handshake (if applicable), and  time spent in TCP slow-start.
Therefore, the abstraction provided by the \vBSDsockets does not assist the
application in distributing traffic among multiple paths. Rather, it puts a huge burden on applications that want to do so.

In conclusion, these problem areas demonstrate that the \vBSDsocketAPI is not well suited to enable multiple access connectivity in an easy and portable way.
In the next Sections \sref{sec:prototype_design} and \sref{sec:prototype_implementation}, we describe how our \prototype overcomes these limitations and provides path- and destination selection as an extension to the \BSDsocketAPI.

\section{Socket Intents Prototype Design} 
\label{sec:prototype_design}

We build a prototype implementation to demonstrate the feasibility of path selection and destination selection as features of the client's OS.
Despite the limitations outlined in \sref{sec:bsdsockets_challenges}, our
\prototype extends \BSDsockets to use \socketintents and moves the selection logic into the OS.
In this section, we first discuss the overall design choices (\sref{sub:prototype_design_choices}) and prototype architecture (\sref{sub:prototype_architecture}).
Then, we describe the designs of the individual prototype components.

\subsection{Design Objectives and Choices} 
\label{sub:prototype_design_choices}

\newcommand{\dobj}[1]{\textbf{\color{MidnightBlue}#1}\xspace} 
\newcommand{\dcho}[1]{\textit{\color{Teal}#1}\xspace} 

The prototype is based on the following \dobj{design objectives} (shown in bold) and \dcho{design choices} (shown in italics).

As we want \dobj{insights about the deployability in today's OSes}, we decide to build our prototype as an \dcho{extension of the \BSDsocketAPI}, which is the base of almost all networking APIs used today.

Since different applications have different requirements, the system should allow them to \dobj{specify their ``intent'' for a given communication unit} as hint for the OS what to optimize for.
Since there is not always a universally ``best'' interface, the system enables \dobj{choosing an appropriate path} for each communication unit, i.e., for each message or connection,
The system also allows \dobj{choosing a destination} for each communication unit from the alternatives provided by name resolution.
To realize these objectives, we provide \dcho{three separate API variants} to
address the trade-offs imposed by integration with the \BSDsocketAPI.

To improve deployability, we chose to explicitly \dobj{not require or suggest changes to the server or the network}.
As a consequence, our means of choosing a path are limited to \dcho{choosing the source address, the interface, and/or the first hop} for each communication unit.

To \dobj{evaluate different selection strategies}, which we call \textsl{policies}, we \dcho{realize them as exchangeable modules}. 
These entities decide which access network to use in a given situation based on the available information.
To \dobj{enable joint optimization across all applications} that use the \prototype, these \dcho{policies are hosted in a central place}.

We want our prototype to \dobj{support path and destination selection for the applications' native communication units}, e.g., an HTTP request.
In today's Internet, these are often not aligned with the communication units of the underlying transport, e.g., TCP streams.
An HTTP-based application will typically try to reuse a TCP stream for multiple HTTP requests to reduce overhead and latency.
This connection reuse logic has to be implemented by each application individually, although it is, typically, not application-specific.
We decided to move this logic into the OS and integrate it with our path and destination selection by providing \dcho{implicit connection pools}.
For each communication unit, our system needs to \dobj{map communication units to the appropriate connection pool} and \dobj{decides whether to reuse an existing connection or to set up a new one}.
As we can only choose a path and destination for new connections, we choose to \dcho{tightly integrate the connection reuse logic} with the path and destination selection logic.

Finally, to \dobj{split large communication units and being able to distribute them over an appropriate set of paths}, we \dcho{enable the use of MPTCP and control of MPTCP's path selection}.


\subsection{Prototype Architecture} 
\label{sub:prototype_architecture}

 \begin{figure}[t!]
     \includegraphics[width=0.8\columnwidth, center, trim=0 0 0 0]{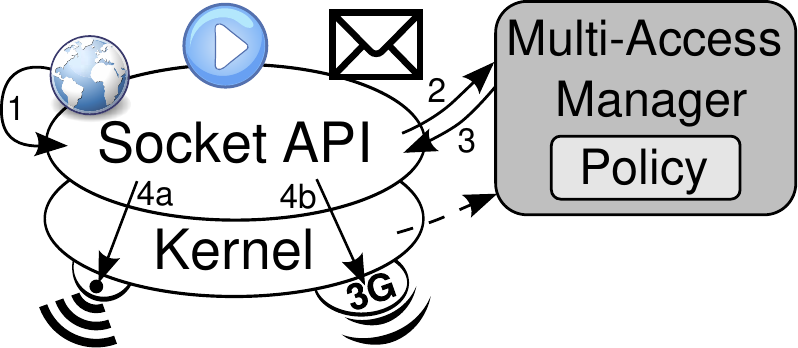}
     \caption{OS Network Stack with Socket Intents.
 	\label{fig:muacc-host-intent}}
 \end{figure}

Our \prototype consists of three components, see~\fref{fig:muacc-host-intent}: Our extended socket API (white), a \polmgr (gray) and a policy modules it hosts (light gray).
In a typical use case, an application specifies its Intents through the API (1), then our
socket library queries the policy within the \polmgr (2).
The policy decides which path(s) to use and communicates this decision back (3), and finally the socket library applies the decision by selecting a path by binding to an appropriate source address (4a/b).


\subsection{APIs} 
\label{sub:prototype_api}

As there is no easy way to integrate application-aware path and destination selection into the \BSDsocketAPI, we provide three different API variants.
While maintaining the UNIX file descriptor as abstraction of a connection, each of these variants takes different trade-offs to overcome the challenges described in \sref{sec:bsdsockets_challenges}:

\begin{itemize}
    \item The \emph{augmented socket-calls} variant, see \sref{sub:muacccontextapi}, follows as close as possible to the call sequence of \BSDsockets.
        It links all socket calls by using an additional context parameter to hold the state needed for path selection beyond the regular parameters.
        It is meant as a baseline to explore which aspects of automated path and destination selection can be integrated into the \vBSDsocketAPI without changing the application logic (besides providing Socket Intents).
    \item The \emph{augmented name resolution} variant performs automated path and destination selection as part of the name resolution, see \sref{sub:getaddrinfoapi}.
        This minimizes the changes to the \BSDsocketAPI, but requires the application logic to change to use the results from path and destination selection.
    \item The \emph{message granularity} variant, see \sref{sub:socketconnectapi}, adds support for access selection at message granularity, e.g., HTTP requests.
        It moves the connection setup into a single API call.
        Thus, it completely replaces the usual call sequence of \BSDsockets in order to enable automated connection caching along with implicit connection pooling.
        This variant is the most powerful one, and therefore used in our evaluation.
\end{itemize}

For all variants, name resolution is off-loaded to the \polmgr and handled by the policy module to work around limitations of the standard resolver library.
See \sref{sec:bsdsockets_challenges} for an extensive discussion about the limitations \BSDsockets impose on implementing automated path and destination selection and their consequences for the implementation of the individual API variants.


\subsection{Policy Design} 
\label{sub:policy_design}

\socketintentpolicies are entities that decide which access network to use for a given communication unit. 
They range from simple static configurations up to complex dynamic algorithms that try to take full advantage of the available information.

To enable informed decisions, these policies need to know about the intents of an application as well as interface parameters and statistics, including byte counters and transport protocol state.
Within its decision logic, the policy needs to respect the optimization of external communication partners, i.e., it should only rely on DNS replies of the same interface (see \sref{sub:dnsifbind}).

For the sake of simplicity, we chose not to support per-application policies, but rely on the information provided by Socket Intents
This also enables us to treat communication units of a single application with different communication needs appropriately.

As a first application-aware policy, we introduce the \eafpolicy:
This policy is based on the idea that downloading objects of different sizes can benefit from different path characteristics, as download time largely depends on the object's file size as well as the RTT and available bandwidth on the path.
We use the \filesizeIntent, which allows an application to hint for the expected size of a communication unit, e.g., allowing an HTTP client to hint about the size of an object to be transferred.
Assuming that there are at least two access networks and they vary in RTT and bandwidth, our intuition is that if the communication unit is small, the policy should choose the interface with the shorter RTT.
If the communication unit is large it should prefer the interface with the larger available bandwidth.
Thus, each unit is scheduled on the interface with the earlier arrival time, resulting in a shorter overall completion time.


\section{Socket Intents Prototype Implementation} 
\label{sec:prototype_implementation}

In this Section, we present the implementation and technical challenges of the three components of our Socket Intents Prototype in more detail: the extended socket APIs, the \polmgr, and the policy modules.
The source code of all components consists of about 15k lines of C and is available under BSD License\footnote{\url{https://github.com/fg-inet/socket-intents/}}.

First, we will explain our extended socket API variants introduced in \sref{sub:prototype_api} in more detail.
All three variants are implemented within a common wrapper library for the \BSDsocketAPI and are portable across Linux and MacOS.
Then, we will dive into the implementation of the \polmgr (\sref{sub:prototype_impl_polmgr}), and the policy (\sref{sub:prototype_impl_policy}).

\subsection{Augmented Socket-Calls API} 
\label{sub:muacccontextapi}

With this API variant, we try to stick as close as possible to the call sequence of \vBSDsockets.
Support for \socketintents is provided by adding an \lname{INTENT} socket option
level and thus allowing to specify \socketintents as socket options.
The challenge of implementing this design is that there is no mechanism to pass state needed for path and destination selection between the \vBSDsocketAPI calls, that is not part of their explicit parameters\footnote{
A kernel based implementation could pass information between all socket calls except \fcall{getaddrinfo} by extending backing struct for file descriptors.}.
To overcome this limitation, we it adds an additional parameter 
to all socket calls including \fcall{getaddrinfo}.

%
%

With this API variant, we can support informed path and destination selection with very few modifications to the application.
However,
due to our way to pass resolver and selection state,
the selection process becomes in-transparent to the application and does not work with applications that want to perform mechanisms like Happy Eyeballs~\cite{rfc6555}.


\subsection{Augmented Name Resolution API} 
\label{sub:getaddrinfoapi}

This API variant realizes path selection and endpoint selection within a modified variant of \fcall{getaddrinfo}.
For all other socket API calls, the application uses the \vBSDsocketAPI.
Socket Intents, alongside with other relevant socket options, are passed directly to our modified \fcall{getaddrinfo} as part of the \lname{hints} parameter.
To do so, we extended the \lname{addrinfo} struct to include a list of socket options and the source address for the outgoing connection.
We also provide a new \lname{socketopt} struct to pass a list of socket options as part of our extended \lname{addrinfo} struct.
The name resolution implementation of \fcall{getaddrinfo} is done by the \polmgr, which makes all decisions and returns them in the \lname{result} parameter as list of endpoints ordered by policy preference.
Each endpoint is annotated with the source address the application should bind to and socket options that should be set on the socket.
Applications use this information as parameter to the \vBSDsocketAPI call or other APIs.
We provide helpers to set all socket options from the \lname{result} data structure on a given socket.

With this API variant, applications can benefit from informed path- and destination selection provided by the \polmgr while maintaining full control over the connection setup, e.g. to perform Happy Eyeballs~\cite{rfc6555}.
This comes at the cost of having to modify the application to bind to a source address and pass socket options through our API.


\subsection{Message Granularity API} 
\label{sub:socketconnectapi}

This API variant offers path and destination selection for message granularity communication units, e.g. HTTP requests, using stream transports, e.g. TCP.
As stated in \sref{sub:prototype_api}, we decided to realize this together with implicit connection pooling to enable connection reuse.
As our focus is on supporting simple request/response type protocols, e.g., HTTP/1.1, we presume sequential reuse of connections by the same application.
We do not yet support multiple concurrent requests on the same TCP connection as in HTTP/2, since this would also require to implement explicit message extraction like \cite{PostSocketsFIT2017} or \cite{linux-kcm} and does not provide us with insights for path- and destination selection.

To expose the functionality, we extended the \BSDsocketAPI by adding three new calls:
\fcall{socket\hspace{0pt}connect} to get a new socket or reuse an existing one, \fcall{socket\hspace{0pt}release} to mark a socket as available for reuse, and \fcall{socketclose} to close a socket (see \tref{table:socketconnect_api}).
Since it realizes the complete connection handling, it moves functionality needed by many applications into the OS.

\begin{table}[tb]
	\centering
	\caption{Message Granularity Socket Intents API}
    \label{table:socketconnect_api}
	\vspace{1pt}
	\scriptsize
    \begin{tabular}{p{2.3cm}p{3.9cm}r}
	    \toprule
	    Call & parameters (excerpt) & in/out\\
	    \midrule
	    \verb!socketconnect! 		& \verb!int *socket! 				& in,out \\
	            					& \verb!const char *host!			& in \\
	            					& \verb!size_t hostlen!				& in \\
	            					& \verb!const char *serv!			& in \\
	    							& \verb!size_t servlen!				& in \\
	            					& \verb!struct socketopt *sockopts!	& in,out \\
	            					& \verb!int domain!					& in \\
	            					& \verb!int type!  					& in \\
	            					& \verb!int proto! 					& in \\
	    \verb!socketrelease! 		& \verb!int socket! 				& in \\
	    \verb!socketclose! 			& \verb!int socket! 				& in \\
	    \bottomrule
     \end{tabular}
\end{table}

When an application wants to send a request, it uses \fcall{socketconnect} to asks our \prototype for a socket for a specific host, service, and socket options (including \socketintents) tuple.
The in/out parameter for the socket file descriptor allows to explicitly request a new socket or reuse one of a set of sockets.
The return value informs the application whether the socket is a new one or an existing one.
This allows the application to decide if it needs to add any per-connection actions, e.g., if a new TLS handshake has to be started for a new connection.
Future work will remove this complication by integrating TLS into the transport stack.
Once the application is done, it can either release or close the file descriptor using the second or third new call.
The former enables reuse, the latter does not.

The \prototype manages an implicit pool of active sockets per destination host/service pair.
The implementation of \fcall{socketconnect} checks whether there are currently unused open sockets to the same host and port.
It then checks with the \polmgr if any of these existing sockets satisfy the needs of the request according to the \socketintents\footnote{
We assume that the policy takes connection setup time into account for this decision.}. 
If not, the \polmgr chooses an interface for a new socket to be created.
The call then either returns the chosen existing socket or the newly opened one.

Using this API variant is very convenient for writing request/response style applications designed around it, but re-writing applications turned out complicated as it totally changes the way connections are handled in an application.


\subsection{Lessons Learned for the APIs} 
\label{sub:impl_api_conclusion}

Adding path and destination selection to BSD Sockets is hard;
Its API calls are not designed to defer choices to a moment where all necessary information is available.
Our API variants address this problem by choosing different trade-offs:

If limiting path and destination selection to the granularity BSD Sockets typically provide today (TCP connections), the Augmented Name Resolution API variant seems to be a good compromise.
Still, it forces applications to change and implement a lot of connection management.

To do path and destination selection at message granularity, we have to make more concessions: We need to add connection caching and pooling to our API.
In student projects using this API variant, we encountered a surprising behavior:  sockets returned by \fcall{socketconnect} returned failures on write.
This happens if the remote side has already closed the connection while it is still in the pool for re-use.
As \vBSDsockets do not provide an explicit mechanism to notify an application about the closure of a socket, we mitigate this behavior by testing the connection before scheduling it for reuse.
This API still uses file descriptors as socket abstraction, but otherwise largely diverges from the \vBSDsocketAPI and requires heavily modifying existing applications.

Besides that, the integration of BSD sockets into file I/O system provided us with many implementation challenges and required many shadow state-keeping and other hacks~\cite{I-D.tiesel-taps-socketintents-bsdsockets}.
Retrospectively, it seems easier to move to a new API like the one we design in the IETF Taps Working Group~\cite{I-D.taps-interface}.


\subsection{\polmgr} 
\label{sub:prototype_impl_polmgr}

The \polmgr runs as a user space service that does not require special kernel support.
It runs as a service available to all applications on the client and is implemented using \emph{libevent}.
Our API uses Unix domain socket to communicate with the \polmgr.
After start up, the \polmgr creates a list of all local interfaces with their network prefixes assigned and loads the policy.

The \polmgr does not keep any per request state, but does not prevent the policy to do so.
A module implementing a policy consists of a set of callback functions which are triggered by our socket library, the \polmgr, and DNS replies.
It can use all functionality provided by  \emph{libevent} as well as the \emph{evdns} resolver library that is pre-configured for each interface  in the \polmgr.

\subsubsection{Estimating Path Characteristics} 
\label{ssub:prototype_impl_ifstats}

To make decisions, the policy uses various network statistics from the \polmgr.
Hereby, the \polmgr periodically queries the OS for smoothed round-trip times (SRTTs) of all current TCP connections to calculate the median and minimum RTT over each available prefix.
Also, based on the interface counters it computes the currently used network bandwidth for each interface.
The query interval is configurable. Our empirical observations suggest that an interval of 100~ms works well. 
The \polmgr also gathers additional passive measurements, such as bit error statistics and current channel utilization as reported by the WiFi access point (if available).
More data about the current network performance can easily be integrated by adding code to the \polmgr or the policy. 
All information is stored on a per interface basis within the \polmgr.


\subsubsection{Resolver Integration} 
\label{ssub:prototype_impl_dns}

The \polmgr has to guarantee that DNS replies are kept separate on a per-interface basis and, therefore, should only be cached and used for communication on the same interface from which they were acquired.
This separation is necessary to avoid interference with DNS-based server selection and load balancing as laid out in \cite{ANRW-multihomed-singlelink}.
As DNS caching is essential for the performance of applications such as Web browsers, we consider this functionality an integral part of the \polmgr and not of the application.


\subsubsection{Controlling Multipath TCP with Socket Intents} 
\label{sub:socket_intents_mptcp}

We use MPTCP to split a TCP stream across multiple paths.
This allows bandwidth aggregation for large transfers and thus complements the per-request scheduling.
As a result, \socketintents can choose appropriate interfaces for both small communication units---which the policy can distribute---as well as large ones---which MPTCP can handle.
In addition, controlling MPTCP from the \socketintentspolicy avoids opening subflows on already crowded interfaces or on interfaces with a high RTT, which can lead to head-of-line blocking for small objects.

To enable the \socketintentspolicy to control the usage of MPTCP we added an additional path manager to the Linux MPTCP implementation.
Our user-space \polmgr uses Netlink~\cite{netlink} sockets to communicate with the kernel-space MPTCP path-manager. If a policy decides to use
MPTCP it selects an interface for the initial subflow.
If MPTCP is feasible the path-manager notifies the \polmgr, so the policy can choose on which interfaces to add subflows.
MPTCP can then distribute the TCP stream over all these interfaces. For more details on our implementation, we refer to~\cite{ma-mirko}.


\subsection{Policy} 
\label{sub:prototype_impl_policy}

The policy implements the logic for deciding which interfaces, and, thus, which source and destination address pair to use.
The actual \socketintentpolicy is implemented as modules for the \polmgr.
It is shared by all applications of a host that use our socket interface, as their individual needs are communicated using \socketintents and are not realized via individual policies.
The policy picks a suitable interface for each communication unit.
Then, if a set of open sockets is given, the policy tries to reuse one of them by selecting a socket which uses the chosen interface from the given set.
If no set is given or if no suitable socket is found, the policy advises the application to open a new connection and suggests an IP address of the chosen interface as source.
When the policy has decided which source and destination address pair to use, it instructs the \polmgr to send this information back to our socket library. With MPTCP, the policy may have to keep track of the requests to aid, e.g., the setup of MPTCP subflows.

We implement the following polices in our prototype:

\subsubsection{Single Interface Policy} 
\label{ssub:sub_prototype_impl_policy_ifx}

This policy always chooses a particular, statically configured interface.


\subsubsection{Round Robin Policy} 
\label{ssub:sub_prototype_impl_policy_rr}

This policy uses multiple interfaces on a round robin basis.


\subsubsection{EAF Policy} 
\label{sub:prototype_impl_policy_eaf}

The \poln{EAF} policy uses the \filesizeIntent to predict the completion time for each
available interface. It then chooses the one where the communication unit
will arrive first.
For \poln{EAF} the \prototype uses estimates of the minimum SRTT per prefix
and the available bandwidth on the interface.
The object size for the \filesizeIntent is determined via a
two-step download that is described in detail in \sref{sec:proxy}.
\poln{EAF} estimates the available bandwidth
by dividing the maximum observed bandwidth on an interface by the number of
already scheduled objects on the same interface.
We divide the file size by the
estimated available bandwidth to approximate the download duration.  We add one
RTT if a connection can be reused and two RTTs if a new connection has to
be established.  Finally, the interface with the shortest predicted arrival
time is chosen.  We do not consider TLS handshakes.


\section{Evaluation Methodology}
\label{sec:methodology}

To evaluate the benefits of \socketintents for Web browsing, 
we use the following methodology

\subsection{Challenges}
\label{sub:methodology-challenges}
\label{sub:methodology-overview}

When evaluating the \prototype to showing performance implications in a generic manner we face several challenges:
First, the evaluations have to be conducted in a realistic environment, including realistic application behavior and network settings.  Second, reproducibility of the gained results has to be assured. Third, we have to evaluate a large number of different application and network settings to gain meaningful results.
We address these challenges as described in the following. 
As application scenario for the evaluation we select Web browsing. To achieve a high degree of realism, we use the Firefox Web browser. Due to the high complexity and optimization level of a browser we decided to implement a client side Socket Intent proxy instead of extending the Web browser. As network settings we decided to rely on typical smartphone settings, i.e., two disjoint paths between the client and the server.

Reproducibility of the results is achieved by excluding most uncontrollable influence factors for the chosen Web browsing scenario. We have to cope with problems like changing Web page content over time, content distribution over various Web pages, and varying backend performance. Accordingly, we crawl different Web pages and mirror them in a local testbed. Thus, most uncontrollable influence factors except for varying execution times of the JavaScript are excluded.  To eliminate the impact of JavaScript we also rely on synthetic workloads without JavaScript. 

To evaluate a large number of different settings we conduct evaluations with around 1000 different Web pages and more than 300 different network configuration. Given the large amount of different settings, we opt for evaluating \socketintents in two ways. 

First, we run a \prototype enabled client-side proxy, which we describe in Section~\ref{sec:proxy}, in a testbed emulating typical network characteristics, see Section~\ref{sub:testbed_setup}.
Second, we run a simulator, which we describe in Section~\ref{sub:simulator-design}, across a wide range of network characteristics and Web pages, see Section~\ref{sub:exp_design}.
To ensure consistency of the results, we validate the runs in the testbed against the simulator, see Section~\ref{sub:validation-proxy-simulator},
and the simulator against the actual download times in the wild, see Section~\ref{sec:validation}.
Finally, our results for the proxy are shown in Section~\ref{sub:proxy_speedups} and for our simulator in Section~\ref{sub:eval_muacc} - \ref{sub:speedup_factors}.

\subsection{Web Proxy}
\label{sec:proxy}
To explore to which extent the possible benefits are viable in practice we
implemented a \socketintent enabled HTTP proxy\footnote{In the future, our \prototype can also be implemented in a Web browser.}.
Our HTTP proxy consists
of 2.300 lines of C~code. Enabling the \prototype took only 20 lines. 

The proxy uses the \filesizeIntent for the objects it downloads, see~\sref{sub:prototype_impl_policy}.
Since the proxy does not know the size of the objects in advance, we
use a two-step download. First, the proxy issues a range request for the first
$m$~bytes\footnote{ Here, $m$ enables a trade-off between RTT and network
  bandwidth.  We see good results for values between 4-8K; values that fit
  within the initial TCP window of today's Web servers.} to get the initial
part as well as the size of the object. If the object is not completely
transferred, the remainder is retrieved via a second range request\footnote{The
  proxy is able to handle various answers including the full object or the
  remaining part of the object, with and without chunked-encoding.}.

\subsection{Web Transfer Simulator}
\label{sec:simulating_web_request_scheduling}

To evaluate the benefits of seamlessly using multiple interfaces and scheduling requests
according to our policies at scale across a wider range of network properties and Web pages, 
we build an event-based data transfer simulator. 
As evaluation metric we use page load time\footnote{Here we focus on network time, i.e., the total time to download the objects of the Web page. 
The complete time to display a Web page also includes times for DNS
resolution, page rendering and possibly client-side JS computation.}, which
has a high influence on the end-user Quality of Experience~\cite{WebQoE}.
Additional metrics can easily be implemented in the simulator, e.g. \cite{kelton17improving}.

\label{sub:simulator-design}

The simulator takes a Web page including all Web objects and their
dependencies (represented as a HAR files --- see \sref{sec:workload}),
the \socketintentpolicy,
and a list of network interfaces with their path characteristics as input. 
The simulator replays the Web page download by transferring all Web page objects
while respecting their inter-dependencies. 
It uses the policy to distribute the object transfers across the interfaces and
calculates the total page load time.

Since our simulator has global knowledge, it knows all object inter-dependencies a priori. Thus, it
can decide when a transfer can be scheduled, i.e, whether all objects that
it depends upon have already been loaded.
To schedule a transfer we assign it to a connection. This is the job of the
policy module which returns either an existing TCP or MPTCP
connection, an interface, or a list of interfaces to use to open a new connection,
or postpones the transfer if the limit of parallel connections has been reached. 
A connection is reused if the host name matches and it is either idle or it
is expected to become idle before a new connection can be established.

The simulator then determines the next event for this connection, such as the completion of a transfer.
When a transfer completes, the simulator records the time, marks all
transfers that depend on it as enabled, and schedules them. 
After the last transfer, the total page load time is reported.

The simulator supports persistent connections with and without pipelining for
TCP as well as MPTCP connections across multiple interfaces. It uses a default
connection timeout of 30~seconds and limits\footnote{These values correspond
to the defaults of the browser we use to retrieve our workload.} the number of parallel connections
per server to 6 and the overall number of connections to 17.
We simulate TCP slow-start using a configurable initial congestion window size with 
a default value of 10~segments~\cite{RFC6928}.
Our motivation for simulating slow-start is to get more realistic load times especially for small objects, which are common on Web pages, when they are downloaded on high latency links, which are common in access networks.
To simulate slow-start and fair bandwidth
sharing, we keep track of the current throughput 
for each connection. 
This is updated according to TCP slow-start and capped by
the congestion window or the available bandwidth share of that interface to
assure TCP fairness\footnote{In our simulator, a connection leaves slow-start once it
reaches the available bandwidth share and never returns to slow-start.}. Our underlying
assumption is that TCP tries to fairly share the available bandwidth between all parallel
connections~\cite{joerg}.
Rather than fully simulating the congestion avoidance of TCP we assume instantaneous
convergence to the appropriate bandwidth share. 
The available bandwidth share of each 
interface is potentially adjusted by each connection event for that
interface. 
If needed the time of the next event is then adjusted accordingly.
Note that for MPTCP, our simulator aggregates the bandwidth of the subflows by simulating them as separate TCP flows. We do not implement coupled congestion control because it does not apply to the network scenario we use in our evaluation, see Section~\ref{sub:network-scenario}.

\subsection{Simulator Implementation} 
\label{sub:simulator-implementation} 

We implemented our data transfer simulator as a heap-based discrete event
simulator. It consists of 3k lines of Python code and is available under a relaxed CRAPL
license\footnote{\url{https://github.com/fg-inet/dtsimulator}}.
It models the process of loading a Web page by keeping track of the status of 
the transfers, connections and interfaces.

Each \textbf{transfer} corresponds to a Web object which contains the object
size, its relationship to other transfers, if the object was transferred via
HTTPS, and the server hostname. The \textbf{connections} are responsible for
estimating and updating the completion times of the transfers which are assigned
to them and for simulating (MP)TCP. In case of MPTCP, we maintain a master
connection and per-interface subflows.  The \textbf{interfaces} bundle the connections and
are used to calculate the available bandwidth shares.

The \textbf{transfer-manager} keeps track of all transfers and informs the
policy if a transfer can be scheduled.  The \textbf{policy} is the main
decision-making entity of the simulation. The policy determines which
interface(s) to use or which connection to re-use for each transfer by
choosing the most appropriate one. The policy then notifies the
transfer-manager to schedule the transfer. 

\label{sub:dependency_web_workload}

To schedule objects in the appropriate order, we derive their interdependencies from the HAR files we gather, see Section~\ref{sec:workload}.
While identifying all objects of a Web page from the HAR files is
straightforward this does not apply to the object dependencies.  Some object
dependencies are obvious from the base page, the HTML document, and the
client-side DOM. However, JavaScript or other Web objects can modify the DOM,
by adding or removing Web page objects, at any point during the page load.  For
example, when a Web site uses JavaScript to dynamically load pictures the
simulator should not start downloading these pictures before the JavaScript
object has been retrieved.  After all, the browser first has to parse the
JavaScript before it can download the pictures.

We decided against using sophisticated systems to derive interdependencies,
e.g.,~\cite{PolarisDOMDependencies}, since their focus is on finding the true
dependency tree to speed up future downloads. Thus, using these dependencies
often leads to much more optimistic results compared to the capabilities of
current browsers. Thus, to ensure compatibility we use a more conservative
heuristic. We identify the dependencies from the download times contained in
the HAR files. This method is feasible since we use a non-bandwidth limited client to
gather the HAR files.

We implement our \eafpolicy, see \ref{sub:policy_design}, for the simulator.
Since the simulator tries to provide an upper bound of the benefits it 
relies on its global knowledge  about all currently active transfers. 
The RTT and maximum interface bandwidth, as well as the size of the objects
for the \filesizeIntent, are known a priori.
Within the simulation, we add one RTT if a connection can
be reused, two RTTs if a new connection has to be established, and
two additional RTTs for each TLS handshake.

Since \socketintentpolicies can use transfer predictions, 
policies can reuse the simulator logic to obtain an estimate of the completion
time given the current state and an interface/connection option. This is
realized by partially cloning the simulator's state, including all currently
active transfers, and simulating the completion time for that transfer.

For \poln{MPTCP}, the simulator presumes that MPTCP subflows can be opened on all local and remote interfaces.
With two network interfaces at the client and one interface at the server,
the policy establishes two subflows. 
The interface for the initial subflow is configurable.
We considered two variants: starting the initial MPTCP subflow on
the same statically chosen interface (\poln{MPTCP if1}) or always on a different, randomly chosen interface (\poln{MPTCP rnd}). 

Finally, \poln{EAF\_MPTCP} combines \poln{EAF} with MPTCP.
In addition to predicting the
arrival time for each interface, it also considers MPTCP for all possible
interface combinations. The intuition here is that MPTCP is beneficial for
some cases but not all cases. For example, this policy can avoid scheduling
small communication units on a high RTT interface.
The simulation of \poln{EAF\_MPTCP} is analogous to the EAF policy, but it includes
predictions with MPTCP for all interface combinations, therefore using the interface for
the first subflow that is predicted to give the best results.
This policy considers neither the \socketintents nor the current network performance.

We test the basic functionality of the
various simulator policies
using various traffic patterns that can take advantage of
\poln{MPTCP}, \poln{EAF}, and \poln{EAF\_MPTCP},
with and without connection reuse.
For these cases we
manually calculate the expected page load times
and check the
simulator results against them.  
In addition,
we use extensive assertions and cross checks within the simulator to ensure the
consistency of the results.

\section{Evaluation Scenario}
\label{sec:scenario}

To evaluate \socketintents using the approach presented in Section~\ref{sub:methodology-overview},
we assume the following network scenario, which serves as the basis for both our testbed setup and our simulator.

\subsection{Network Scenario}
\label{sub:network-scenario}

\begin{figure}[t!]
  \begin{center}
    \includegraphics[width=0.8\columnwidth, trim=0 0 0 0]{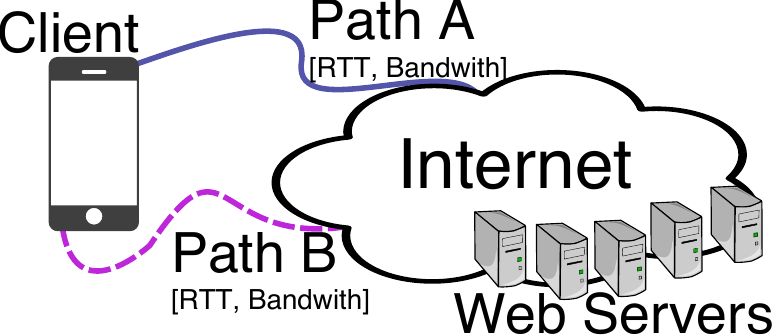}
    \caption{Simplified Network Scenario.}
    \label{fig:network_scenario}   \end{center} \end{figure}

The motivation for our network scenario is that, for mobile devices, access
networks almost always dominate performance as the bandwidth bottleneck is likely to be located there and access networks often introduce major delays\footnote{Internet backbone delays are in the order of a few milliseconds while access delays are typically significantly larger. With increasing access network capacities, the bottleneck might be in the core in some cases. However, capacities are not increasing in all regions of the world.}.
Thus, our network scenario, see \fref{fig:network_scenario}, consists of a
client, Web servers, and the paths between them. We presume that all Web
servers are reachable via both network paths.  Moreover, we choose to neglect
the RTT variability introduced by the Internet since queuing delays on Internet
core links ($\geq$10\,Gbit/s bandwidth) are
negligible~\cite{sundaresan2011broadband}. Therefore, we capture the
path characteristics as ``interface'' RTT and bandwidth. To model connection
reuse, we assume a separate server per hostname.

\subsection{Testbed Setup}
\label{sub:testbed_setup}

\begin{figure}[t!]
  \includegraphics[width=0.8\columnwidth, center, clip=true, trim=0 0 0  0]{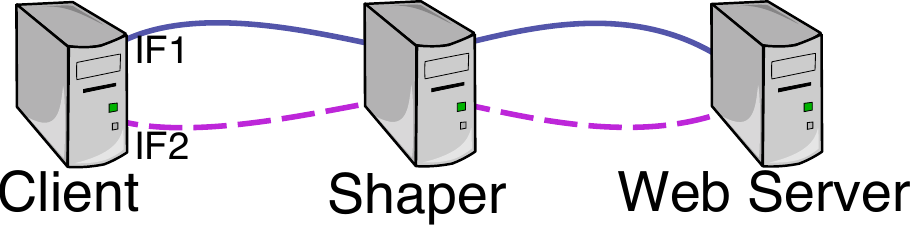}
 \caption{Testbed setup used in the emulation. \label{fig:testbed_setup}}
\end{figure}

To evaluate the benefit of \socketintents in the Web proxy using the \prototype under different access network characteristics
we setup a testbed according to our network scenario, see \fref{fig:testbed_setup}. It consists of three
physical machines: Web server, traffic shaper, and Web client.
The Web client has two network paths to the Web server via two separate network
interfaces. The network characteristics are emulated by the traffic shaper and
include three scenarios of access network characteristics. The characteristics range from fully symmetric to asymmetric,
see~\tref{tab:proxy_scenarios}, and are representative of access network characteristics found in literature\cite{sommers2012cellVSwifi}.
On the Web client, we run a Web browser along with the proxy and the \polmgr.
Our \polmgr supports \poln{EAF} as discussed in \sref{sub:prototype_impl_policy}, which we compare against the use of a single interface.

We automate a browser\footnote{We use Mozilla Firefox~52.5 Web browser with the Selenium browser
automation framework.}, which we
restart for each measurement to ensure that the cache is cold.  
We run the DNS server on the Web client to ensure that name resolution does not add delays.
Furthermore, since the two-step download does not work with HTTPS, we use HTTP.

The Web server hosts our workload. It consists of handcrafted pages,
each with a different number of objects (ranging from 2 to 64) of various
sizes (between 1~KB and 1~MB), as well as mirrored versions of several Web pages
from the Web workload, see \sref{sec:workload}.
We select a range of Web pages that represent our workload in terms of numbers and size of objects and hosts.

We use a single Web server since we assume that Web
performance is dominated by the different access networks.  However,
we set up a virtual host per host name to restrict connection reuse
appropriately. As we want a lower bound of the performance benefits, we tune
the TCP parameters of the Web server to conservative values\footnote{
We use TCP/Reno with an initial congestion window size of 10~MSS. We disable
TCP metrics saving to prevent congestion window caching as well as TCP
segmentation off-loading to eliminate interference with the NIC firmware.}.
On the shaper we emulate a given maximum bitrate using tc and latency with
netem and we restrict queue sizes to avoid buffer bloat. 

\begin{table}[t]
	\caption{Testbed shaper: Network parameters.
	    \label{tab:proxy_scenarios}}
    \vspace{1pt}
	\centering
	\resizebox{0.95\columnwidth}{!}{

\begin{tabular}{ l| r r@{.}l r@{.}l | r r@{.}l r@{.}l }
\toprule
                                 &  \multicolumn{5}{l|}{\bf Interface 1}                                                                      &  \multicolumn{5}{l}{\bf Interface 2} \\
                                 &  \multicolumn{1}{l}{\bf RTT}    & \multicolumn{2}{l}{\bf Down     }   & \multicolumn{2}{l|}{\bf Up}       &  \multicolumn{1}{l}{\bf RTT}    & \multicolumn{2}{l}{\bf Down}  & \multicolumn{2}{l}{\bf Up} \\
                                 &  \multicolumn{1}{l}{ms}   	   & \multicolumn{2}{l}{MBit/s}    	     & \multicolumn{2}{l|}{MBit/s}       &  \multicolumn{1}{l}{ms}         & \multicolumn{2}{l}{MBit/s}  & \multicolumn{2}{l}{MBit/s}\hspace{-1ex}\\
\midrule
\hspace{-1ex}{\bf Symmetric    } &   45                            &    10 &  0                           &    1 &  0                        &   45                            &    10 &  0                         &     1 & 0           \\
\hspace{-1ex}{\bf Asymmetric   } &   20                            &     6 &  0                           &    0 &  768                      &   70                            &    13 &  0                         &     6 & 0           \\    
\hspace{-1ex}{\bf Highly Asym. } &   10                            &     3 &  0                           &    0 &  768                      &  100                            &    20 &  0                         &     5 & 0           \\
\bottomrule
\end{tabular}

}
\end{table}

\subsection{Experimental Design for Simulator Evaluation}
\label{sub:exp_design}

To evaluate the potential benefits of using \socketintents across a wide range of parameters, 
i.e., with different policies under different network scenarios and for different Web pages,
we use a full factorial experimental design.
Each factor can, in principle, influence the page load time. For each factor, we consider multiple values that cover the possible value ranges.  By simulating all
combinations, see \tref{tab:simulation_levels}, we run 9M simulations.

\begin{table}[b!]
\caption{Levels of the Factorial Experimental Design.}
\label{tab:simulation_levels}
\vspace{1pt}
\centering

\begin{tabular}{lp{5cm}}
\toprule
{\bf Factor}              &  {\bf Levels} \\
\midrule
{Policy:                } & \poln{Interface 1}, \poln{Interface 2} \\ 
                          & \poln{Round Robin} (starting on if 1), \\
                          & MPTCP starting on Interface 1 (\poln{MPTCP if1}) or on a random interface (\poln{MPTCP rnd}), \\
                          & Earliest Arrival First (\poln{EAF}), or\\
                          & EAF with MPTCP (\poln{EAF\_MPTCP}).\\
\midrule
{Web page:              } & Alexa Top~100 and Top~1000. \\
\midrule
{Interface 1 RTT:       } & 10, 20, 30, or 50~ms.\\
{Interface 1 Bandwidth: } & 0.5, 2, 6, 12, 20, 50~Mbit/s.\\
{Interface 2 RTT:       } & 20, 50, 100, or 200~ms.\\
{Interface 2 Bandwidth: } & 0.5, 5, 20, or 50~Mbit/s.\\
\bottomrule
\end{tabular}
\end{table}

In our experimental design, the primary factor is the \textbf{Policy} used with all of
our \socketintentpolicies, see~\sref{sub:simulator-implementation}, as levels.
The \textbf{Web pages} of our workloads, see~\sref{sec:workload}, are the second factor: Here, the levels are the different Web pages (with their 26 repeated crawls for Alexa Top~100 and one crawl for the Alexa Top~1000).

The remaining four factors describe the network scenario: Since our simplified network
scenario as illustrated in \fref{fig:network_scenario} consists of one client using
two access networks and various Web servers which are reachable via both interfaces,
these factors are: \textbf{Interface~1 RTT} and \textbf{Bandwidth} as well as 
\textbf{Interface~2 RTT} and \textbf{Bandwidth}.
The levels for these were chosen to reflect typical interface characteristics: 
We consider mobile devices that have WiFi as well as cellular connectivity.
Interface~1 should resemble the possible characteristics of home broadband connectivity
(e.g., DSL or cable) and Interface~2 should resemble the range of possible 3G/LTE
coverages\footnote{Costs or restrictions of the data plan are beyond the scope of this
paper, but could easily be taken into account by an elaborate policy.}.
This results in the levels shown in Table~\ref{tab:simulation_levels}.

\subsection{Web Workload}
\label{sec:workload}

To get a wide range of Web pages
we crawl the landing pages of the Alexa
Top~100 Web sites on 26~consecutive days starting on December~07
2015 and the Alexa Top~1000 Web sites on October~10 2016\footnote{
 \url{http://www.alexa.com/topsites}}.
We focus on the mobile version of the pages by 
overriding the user-agent of our Firefox browser, 
impersonating a generic Android mobile device.

As browser we use Firefox version 38.4.0 automated with Selenium and the Firebug~2.0.13 and NetExport~0.9b7 plugins
to retrieve the objects and to record the crawled Web pages
in the HTTP Archive (HAR) format. Each HAR file contains a summary of all
objects of the page as well as their sizes, types, origins (remote sites), and
timings.  
We use a single vantage point with a high available
network bandwidth, a virtual machine within a university network. 

While most of the pages comprise between 1 and 50 objects there are some with
more than 100 objects or even up to 260 objects. Moreover, many Web pages have
a low median object size. Furthermore, the number of hosts that have to be
contacted ranges from a single one to more than 20 with a median of 7.
The total size of the Web pages is 
between 23.1~KB (5th quantile) and 1.8MB (95th quantile) with a large fraction
of pages below 300~KB.  
These results are in line with previous
work~\cite{ihm2011towards,butkiewicz2011understanding}.

For comparison with less complex Web workload, we add handcrafted Web pages to our workload.
These pages consist of different number of objects (ranging from 2 to 64) of various
sizes (1~KB to 1~MB), and a mix of these objects.

\section{Evaluation}
\label{sec:eval}

To explore the benefits of seamlessly combining multiple access networks for speeding
up Web page load time, we evaluate \socketintents in two ways, see Section~\ref{sec:methodology}.


\subsection{Socket Intent Benefits in Testbed}
\label{sub:proxy_speedups}

\begin{figure}[t!]
    \includegraphics[width=\columnwidth, right, clip=true, trim=0 0 0 0]{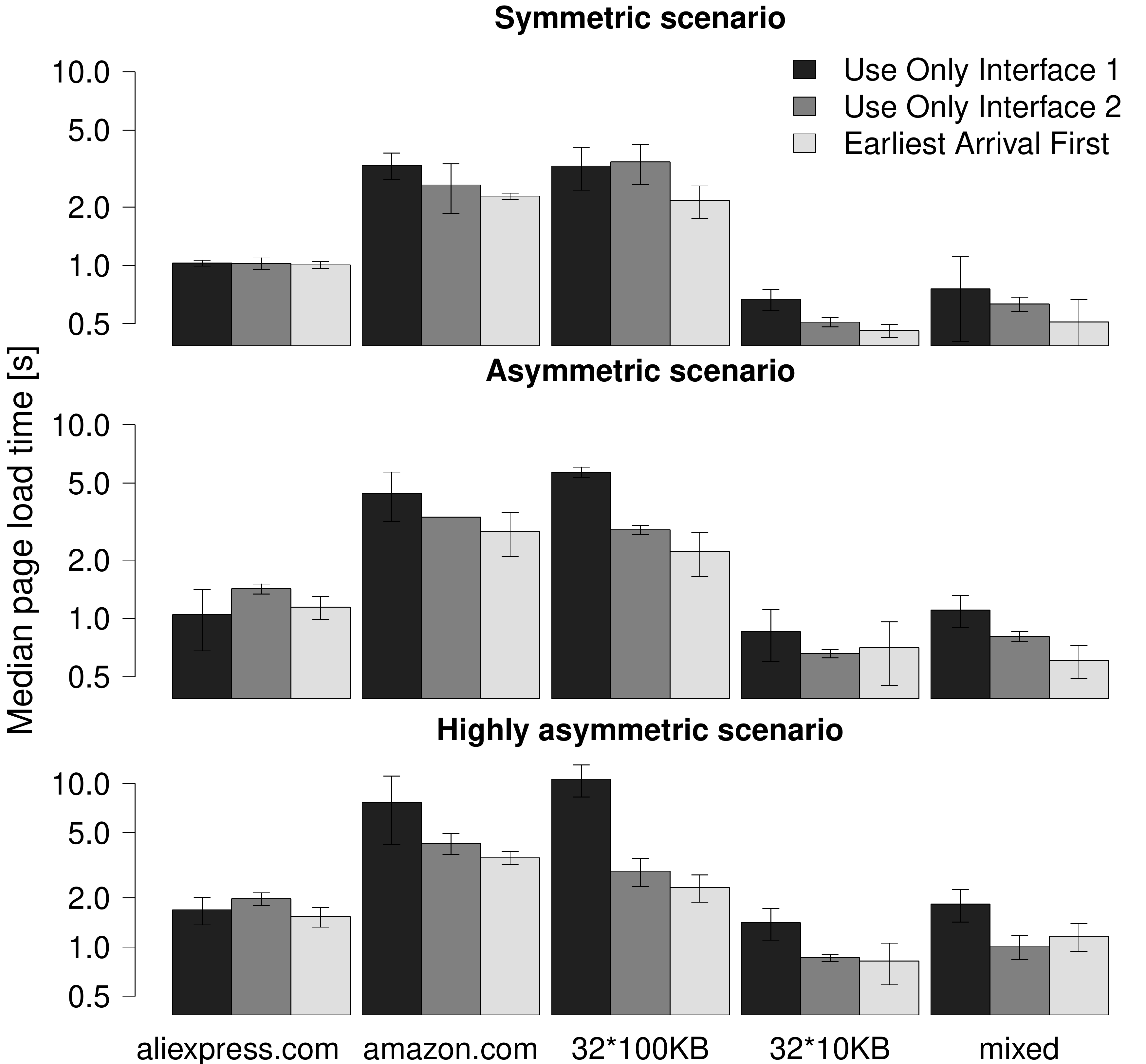}
	 \caption{Proxy: Page load times.\label{fig:barplots}}
\end{figure}

We setup our testbed as described in~\ref{sub:testbed_setup}.
For each of our Web pages we repeatedly download each page 7
times, using a single interface as well as using \poln{EAF}, see Section~\ref{sub:prototype_impl_policy}.
We compute the load time of the individual objects, and aggregate the times
during which objects were downloaded to compute the total page load
time\footnote{The total display time includes page rendering and client-side
JavaScript computation, which we exclude here.}. 
The resulting page load times are shown in
Figure~\ref{fig:barplots} using a logarithmic y-axis. It includes all three
policies: \poln{Interface~1}, \poln{Interface~2}, and \poln{EAF}.
The mixed handcrafted workload shown here consists of 16 objects of 1KB, 8
objects of 10KB, and 4 objects of 100KB.

We
see that \poln{Interface~1} is the better choice if the Web objects are small
and the network scenario is asymmetric. \poln{Interface~2} is the better choice if the
objects are larger or if there are more objects. 
Using both interfaces is, in
particular, beneficial for the symmetric scenario. While there is still a
benefit of using both interfaces it gets smaller for more asymmetric scenarios.

The \poln{EAF} policy takes advantage of the multiple access networks seamlessly.
It either uses both interfaces or the better one of the two with only a
slight increase in page load time variability.
For pages with many objects such as the handcrafted workload of 32 objects of 100 KB, 
our \eafpolicy outperforms the better of the two interfaces with speedups
from 25\% to 50\%.
For some of the actual Web pages that we mirror on our testbed, including \url{amazon.com},
we get speedups in the range of 20--45\%.
For other Web pages such as \url{aliexpress.com}, we only get a speedup of up to 10\%.

The reason for the ``decreased'' benefits compared to the handcrafted pages are that the
mirrored Web pages
fetch content from different servers, which limits connection reuse.
Furthermore, even for mirrored versions of the same Web page, load times vary based on
optimizations in the contained Javascripts, as the Alexa~100 pages are heavily optimized.

Nevertheless, our results highlight the potential of \socketintents: Even with a proxy
the page load times improve. Including \socketintents within the browser rather than a proxy is likely
to yield even better performance.

\subsection{Validation of Proxy and Simulator}
\label{sub:validation-proxy-simulator}

\begin{figure*}[ht!]
  \begin{center}
     \subfloat[Single Interface (50ms, 6Mbit/s)] {
      \includegraphics[width=0.65\columnwidth, clip=true, trim=0 0 0 0]{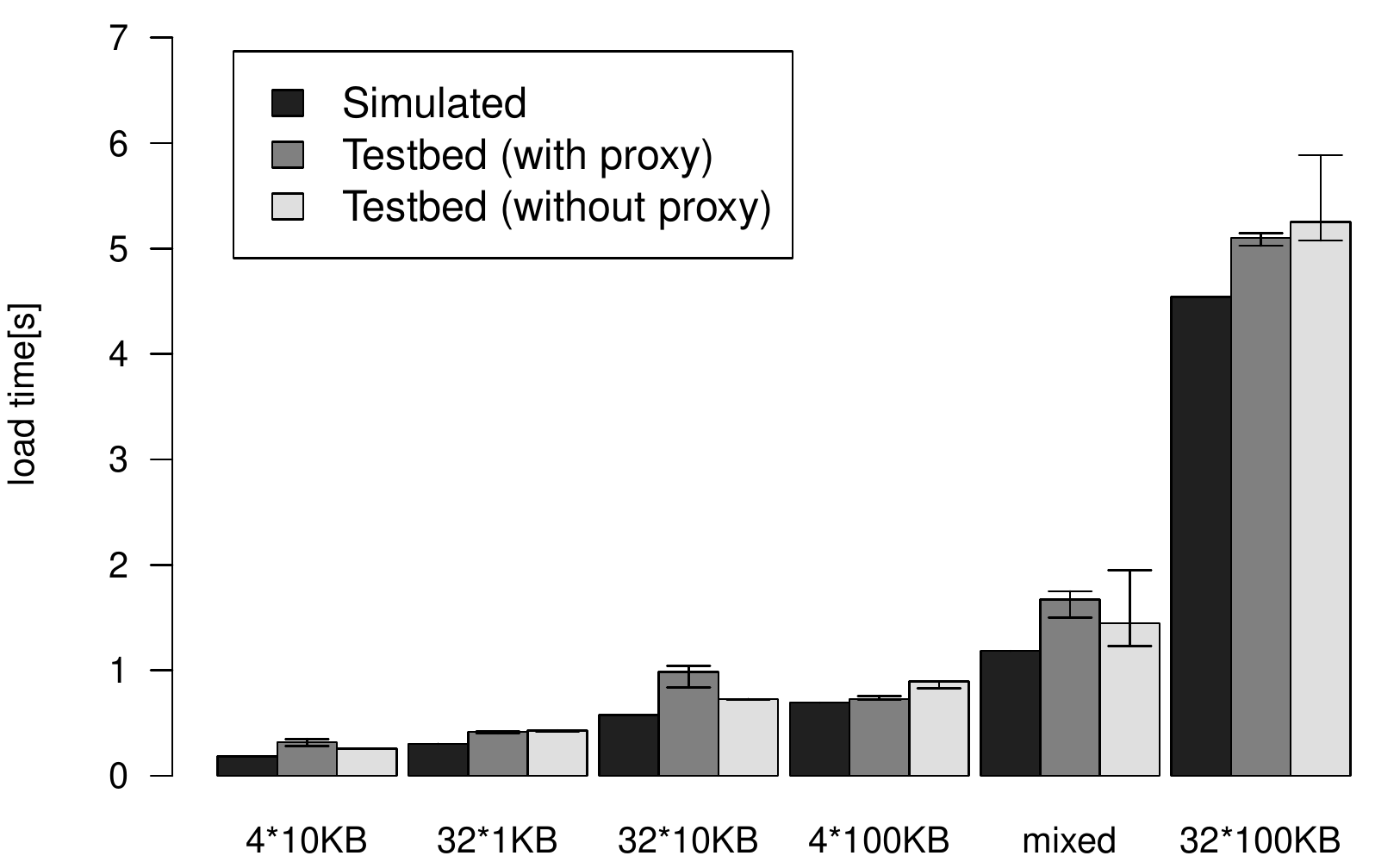}
     \label{fig:validation-50ms-6mbit}
     }
     \subfloat[Single Interface (10ms, 0.5Mbit/s)] {
      \includegraphics[width=0.65\columnwidth, clip=true, trim=0 0 0 0]{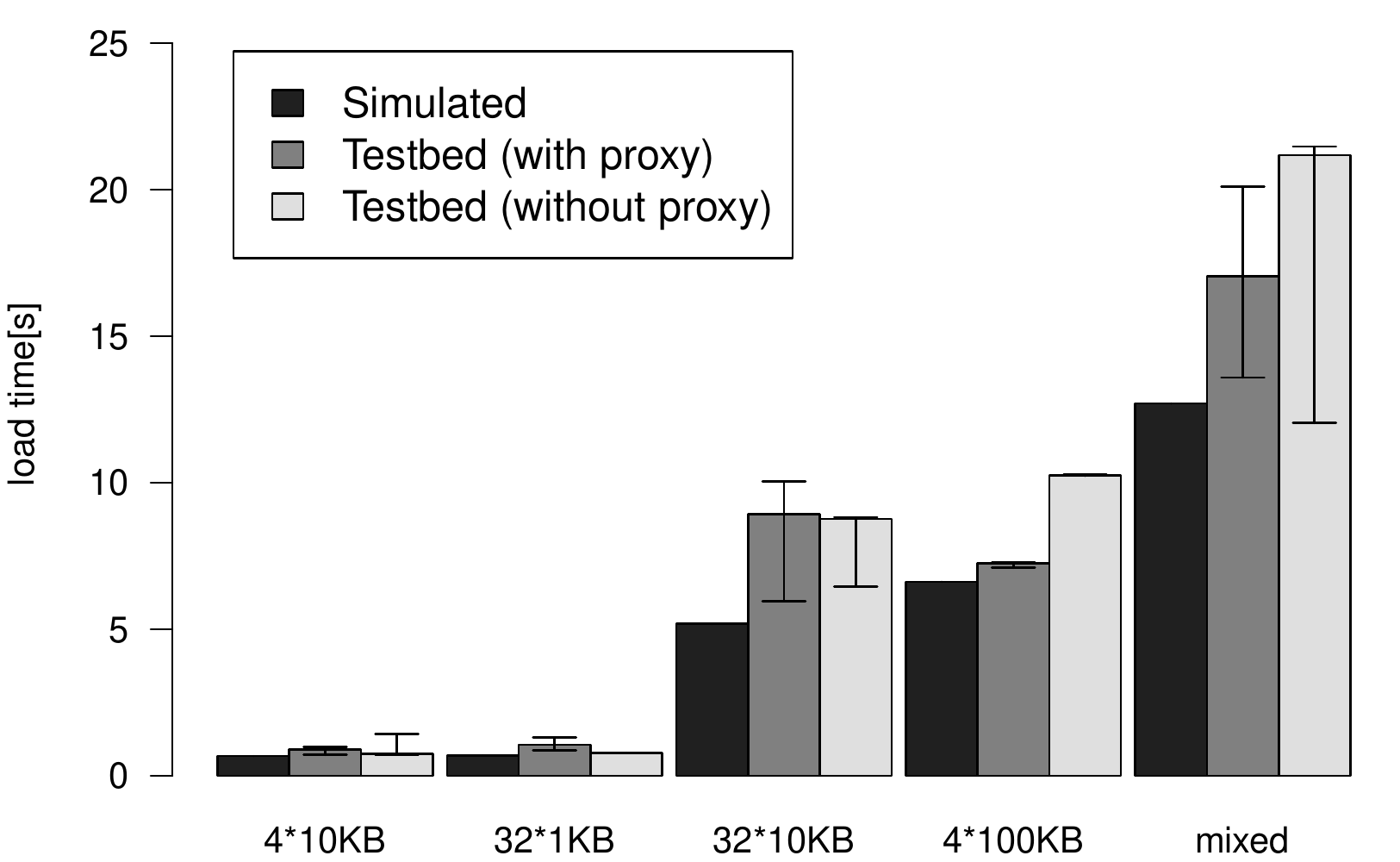}
       \label{fig:validation-10ms-0.5mbit}
     }
     \subfloat[EAF with symmetric shaping] {
      \includegraphics[width=0.65\columnwidth, clip=true, trim=0 0 0 0]{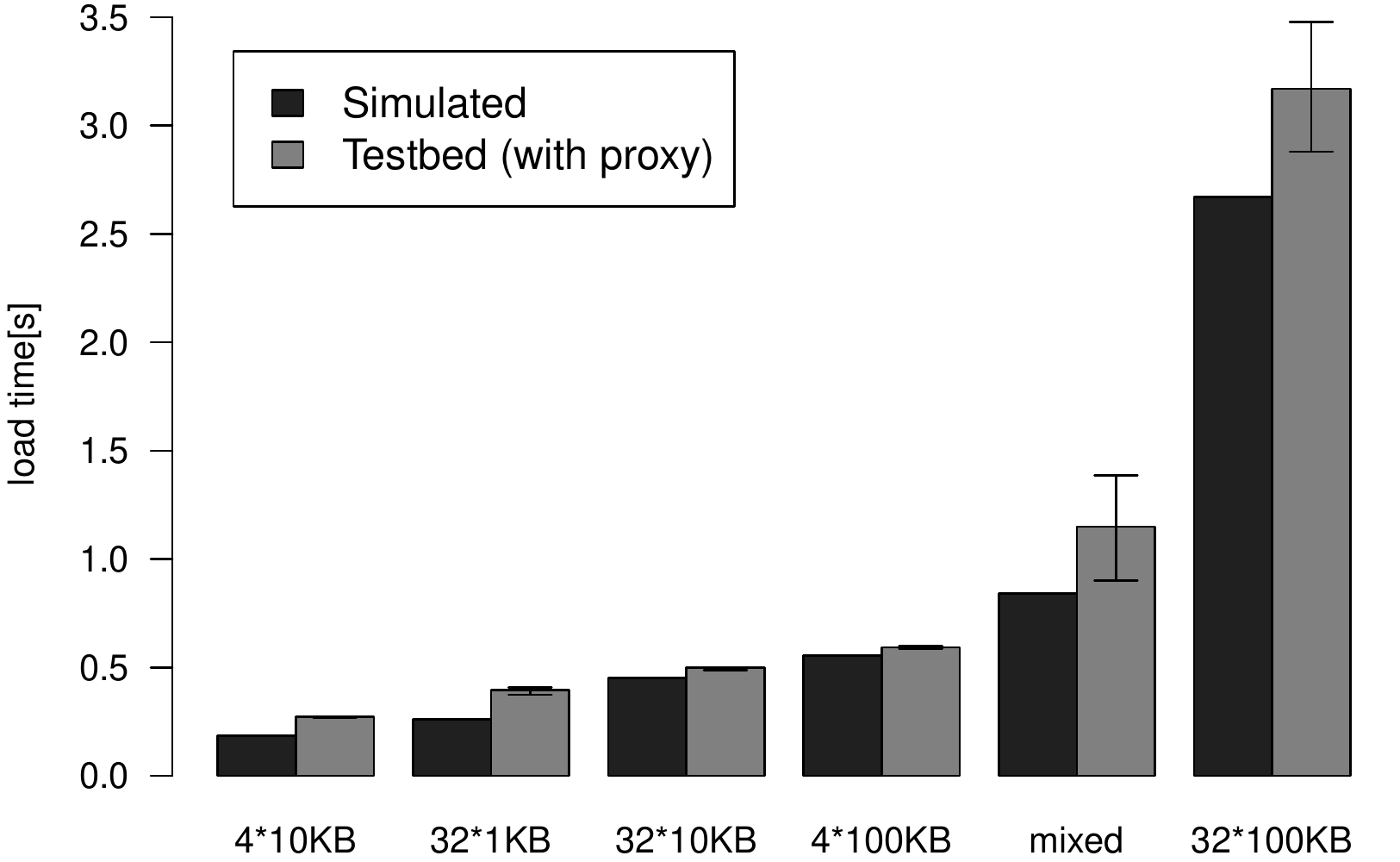}
       \label{fig:validation-eaf-sym}
    }
 \caption{Comparison of simulated load time and actual load time in the testbed with different synthetic workloads.\label{fig:validation_proxy_sim}}
 \end{center}
\end{figure*}

We validate our simulation results against the proxy by measuring the Web page
load time of our workload in the testbed with similar shaper settings as the interface parameters we use in the simulator.
In Figure~\ref{fig:validation_proxy_sim} we compare the simulated and the actual load times for the handcrafted workloads of different sizes, showing the median load time and the 95\% confidence intervals. The mixed workload consists of 32 objects of 1KB, 16 objects of 10 KB, 2 objects of 100 KB and 2 objects of 200 KB.
Using a single interface with an RTT of 50 ms and a bandwidth of 6 Mbit/s, see Figure~\ref{fig:validation-50ms-6mbit}, we see slightly higher load times on the testbed both with and without the proxy, especially for large workloads.
Using a single interface with only 0.5 Mbit/s, see
Figure~\ref{fig:validation-10ms-0.5mbit}, we do not get a page load time for
the workload with 32 objects of 100 KB because the browser times out after 10-20 seconds, so we do not show it in this plot. Using our
\eafpolicy with symmetric shaping (50 ms and 6 Mbit/s on one interface, 50ms and 5 Mbit/s on the other), we cannot test the case without proxy, as we cannot use \poln{EAF} without the proxy. Both our simulator and the proxy in the testbed show speedups, see
Figure~\ref{fig:validation-eaf-sym}. Note the differences between the y axes, which reflect the speedups observed in Section~\ref{sub:proxy_speedups}.
We get similar results for RTTs up to 200 ms and bandwidths up to 50 Mbit/s.

We get similar load times with and without the proxy.
This shows that the two step download in our proxy does not have a major influence on the load time.
Overall the simulator is more optimistic than the testbed. However, the differences are quite small.
The differences to the simulator can be explained by the following observations: 
First, 
the gzip transfer encoding
conflicts with range requests: Sometimes the server sends the whole object even
though only the initial part is requested. Moreover, disabling compression for the initial request is
not feasible as it eliminates compression also for the second request since the
content-range refers to the range after compression.
Second, the simulator
presumes that all independent transfers start immediately, which is not always
the case in practice. This can skew timings, in particular for small
workloads.
These effects are independent from the use of our \prototype.
Accordingly, we can use the simulator to conduct a realistic comparison between scenarios with and without \socketintents.

\subsection{Simulator vs.\ Actual Page Load Time}

\label{sec:validation}

\begin{figure}[t]
  \includegraphics[width=0.95\columnwidth, center, clip=true, trim=0 0 0 0]{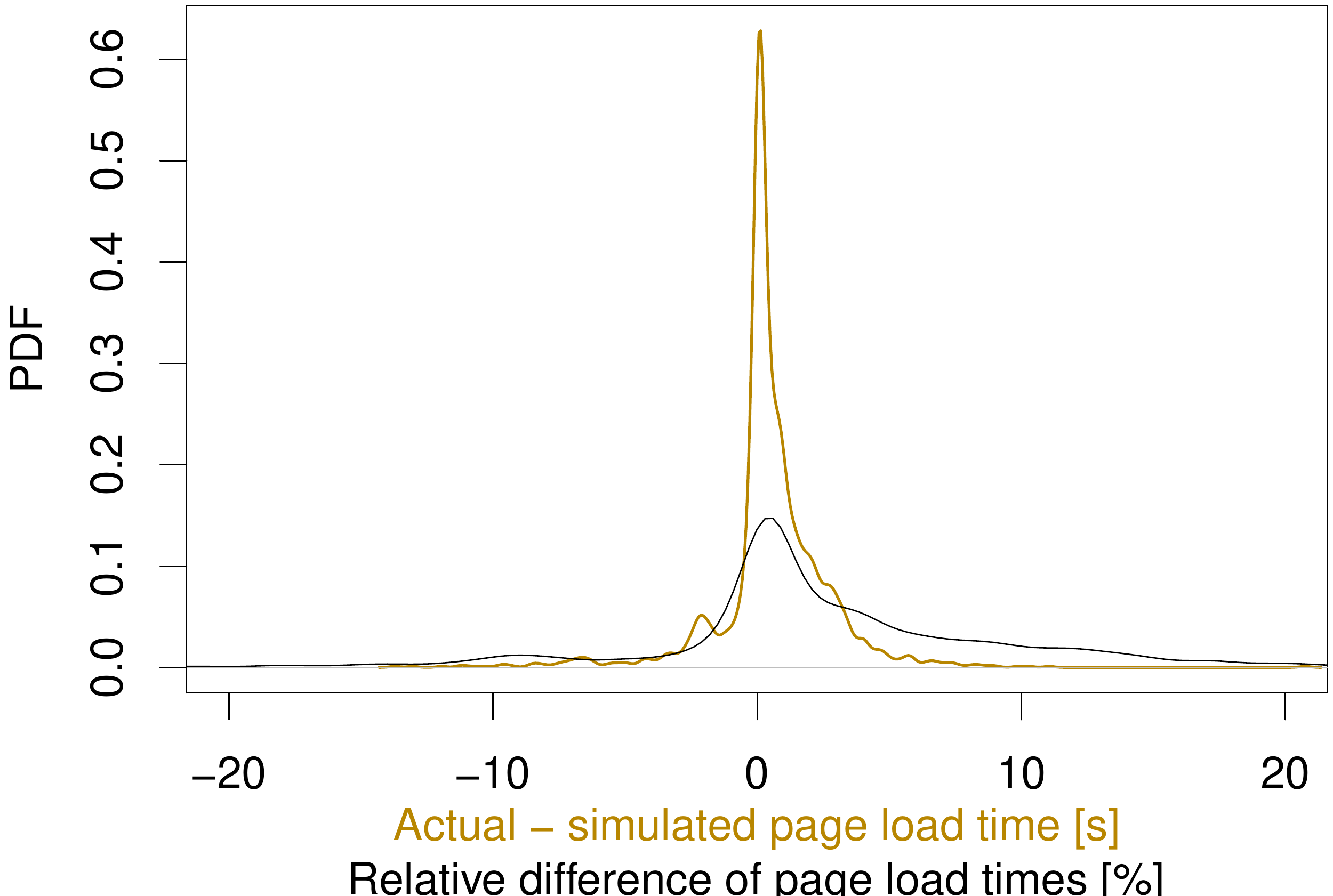}
 \caption{Simulator validation: Relative and absolute difference of simulated
   vs.\ actual page load time. \label{fig:validation}}
\end{figure}

We compare the actual page load time to the simulated one for all Web pages of our
workload.
Given that our crawl uses a machine with a single interface we also use
a single interface with the policy ``Single Interface''. To determine the
interface parameters we estimate the available bandwidth as well as the RTT to
the servers from the actual download. To estimate the available bandwidth we
use all objects larger than a minimum size of 50~KB. Hereby, we take into
account that several of these can occur in 
parallel. Using the median of the estimated bandwidth results in a
typically used bandwidth of 67.13~Mbit/s -- this suggests that none of the transfers
were actually bandwidth bound. To estimate the RTT the simulator
issues a series of pings for each Web page. The median RTT of all servers
of that Web page is then used as an estimator for the interface for the
validation run for that Web page.

The simulator as well as the validation uses several simplifications. First,
the simulator assumes that all Web objects share a single network
bottleneck and that the RTT is the same for all servers.
In reality, some embedded objects of Web pages are fetched from hosts
with different network bottlenecks and RTTs. We use ICMP ping rather than TCP
ping and the pings are not executed while the HAR files are gathered.


Figure~\ref{fig:validation} shows the absolute as well as the relative
differences of the simulated vs.\ the actual page load times for all Alexa
Top~100 Web pages from~\sref{sec:workload}.  The main mass of
both distributions is around zero indicating that the simulated page load times
are very close to the actual ones. This is confirmed by the median value which
is 0.3548/1.5\% for the absolute/relative differences. This highlights that the
simplifying assumptions of the simulator still enable us to approximate the
actual page load times and that we capture most of the intra Web page
dependencies.

There are some differences for some Web pages. We manually checked them and
find a majority is caused by differences in the estimated
bandwidth, server delays, and name resolution overhead. These are, e.g., related to
Web back-office interactions~\cite{backoffice}. Overall, the results are rather
close and show that our simulations result in reasonable approximations
of the actual Web page load time.

\subsection{Benefits of Combining Multiple Access Networks}
\label{sub:eval_muacc}

\begin{figure*}[t!]
    \subfloat[Full Data Range] {
    \includegraphics[width=\columnwidth, clip=true, trim=0 20pt 0 54pt]{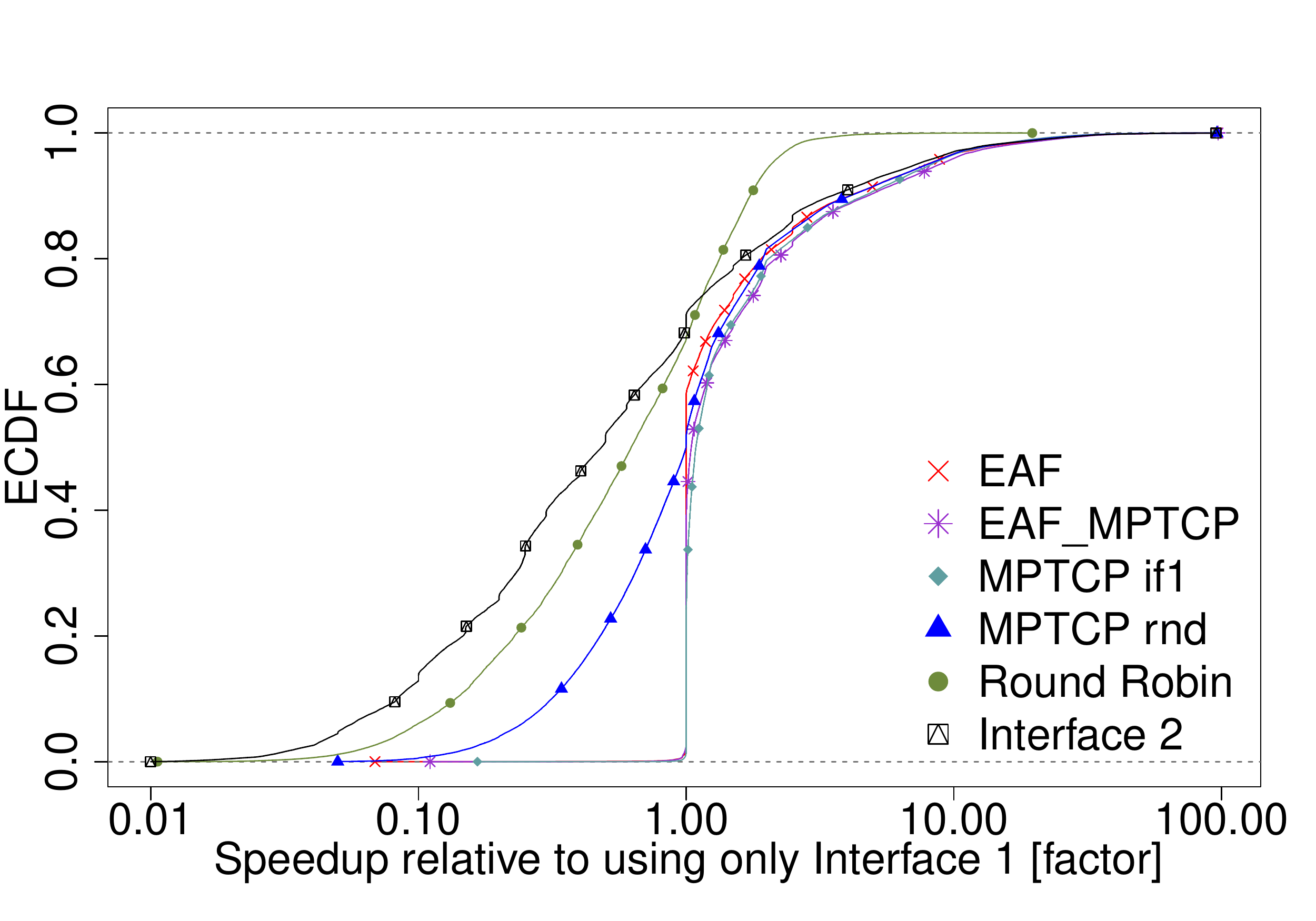}
    \label{fig:speedup_cdf_100}
    }%
    \vspace{\tabcolsep}
    \subfloat[Speedups between 1 and 5.] {
    \includegraphics[width=\columnwidth, clip=true, trim=0 20pt 0 54pt]{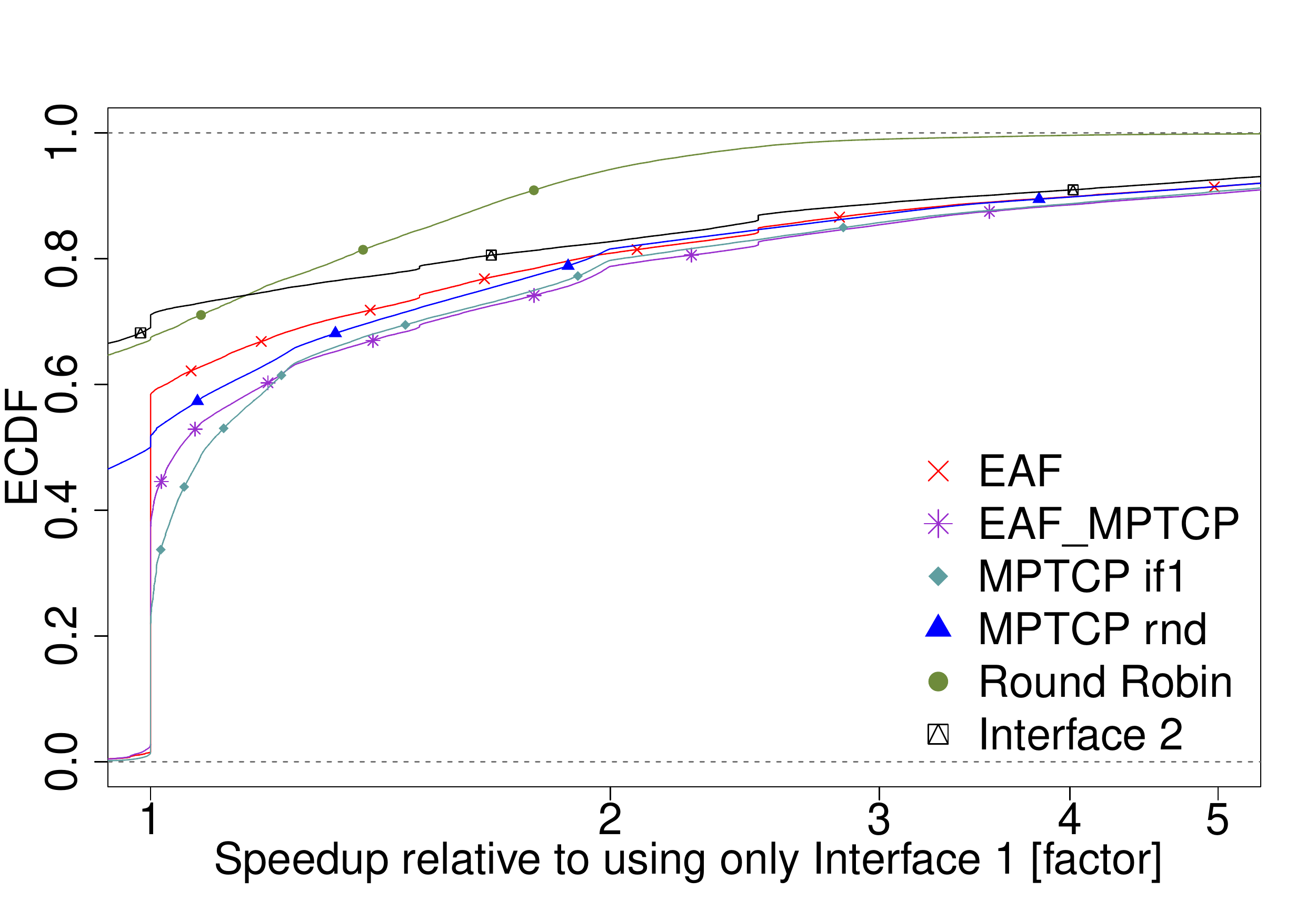}
    \label{fig:speedup_cdf_100_zoom}
    }
 \caption{CDF of Speedups vs. \poln{Interface~1}for the Alexa Top 100 workload.\label{fig:speedup_cdf_100_both}}
\end{figure*}

\begin{figure}[t!]
    \includegraphics[width=\columnwidth, clip=true, trim=0 20pt 0 54pt]{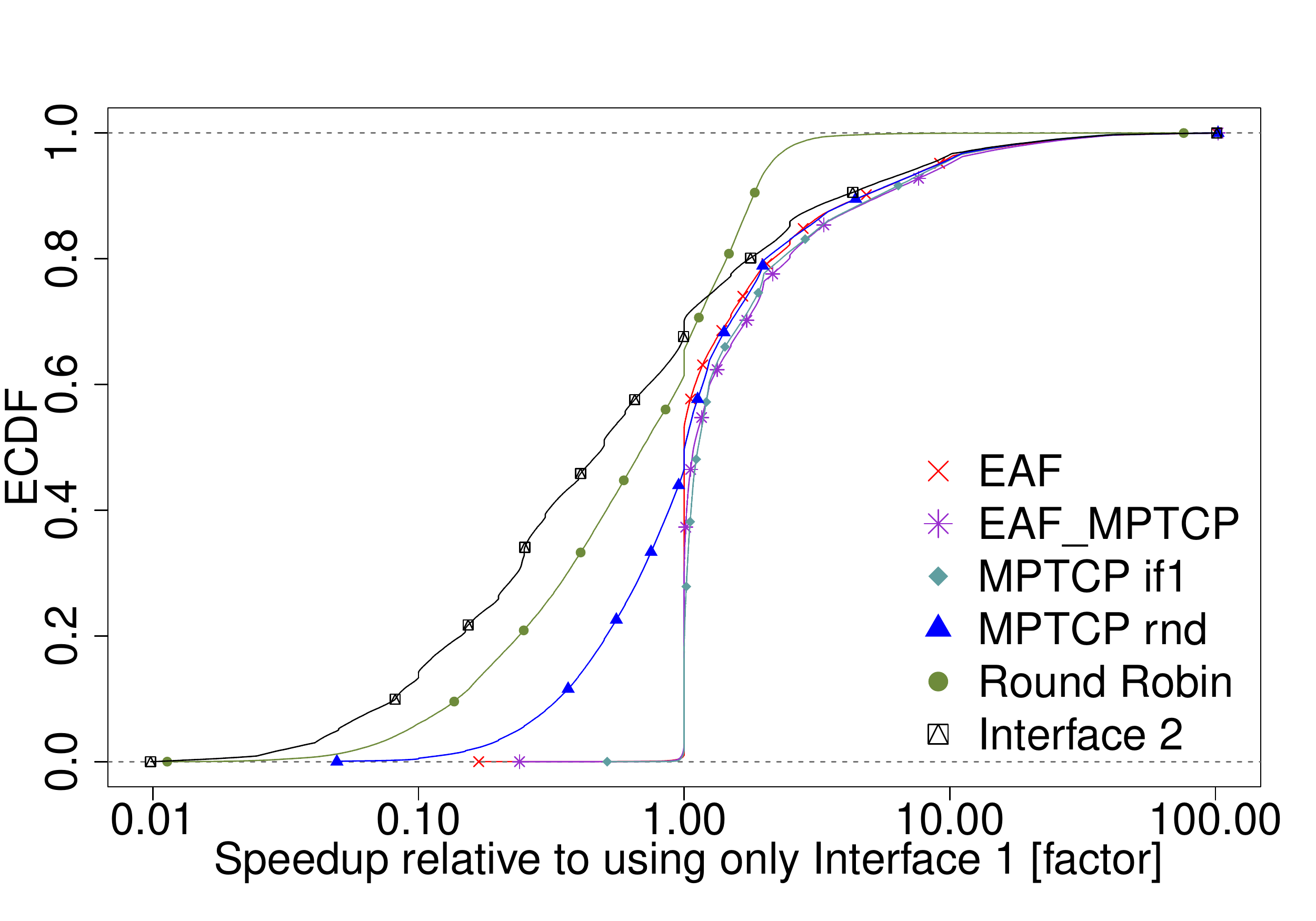}
  \caption{CDF of Speedups vs. \poln{Interface~1} for the Alexa Top 1000 workload. \label{fig:speedup_cdf_1000}}
\end{figure}

To explore the benefits of combining multiple access networks by using
\socketintents, we compare the speedups of the page load times against the baseline
policy \poln{Interface 1}. The baseline policy \poln{Interface 1} resembles
what most current
mobile OSes do: Use only WiFi and, therefore, the home broadband if available.

\fref{fig:speedup_cdf_100} shows the empirical cumulative distribution functions
(ECDF) of the speedups achieved using a simulated \socketintentpolicy relative to only using \poln{Interface 1}, all other parameters being equal.
Thus the ECDF shows speedups across all network scenarios outlined
in \tref{tab:simulation_levels} based on the Alexa Top~100 Web pages and categorized by
the \socketintentpolicy used.
We see that in more than 42\% of the cases for \poln{EAF} and 63\% of the cases for
\poln{EAF\_MPTCP} these policies provide a speedup of more than 1, which means that
loading a Web page using these policies is faster than using \poln{Interface 1} in the
same scenario.
In the remaining cases, they almost always provide a speedup of 1, which means that they
neither gain nor lose from using multiple interfaces.
In these cases, the page load was not bandwidth limited and simply loading 
the page over Interface~1 was the fastest option. Thus using the
other interface in addition did not provide any speedup.
Therefore, \poln{EAF} and \poln{EAF\_MPTCP} simply choose to use Interface 1 in these
cases.
We also see that in about 1.5\% of cases \poln{EAF} and \poln{EAF\_MPTCP} is slower than  
\poln{Interface 1} which turned out to be a limitation of the simulator\footnote{
In these cases, the simulator fetches a single huge object via the less suitable
interface while the connection limit prevents starting a new connection on the more
suitable one.}.
Overall these results show that using \poln{EAF} and \poln{EAF\_MPTCP} is a good choice in any case.

The speedups of both MPTCP policies are very dissimilar: When establishing the first subflow over Interface 1 (\poln{MPTCP if1}), it shows a speedup greater than 1 in 78\% of the
cases and neither improvement nor penalty in the other cases. In contrast, if starting the first subflow for MPTCP over a randomly chosen interface (\poln{MPTCP rnd}), MPTCP performs worse than
\poln{Interface 1} in 48\% of the cases and can be up to 10x slower.
We take a closer look at these effects in \sref{sub:eval_mptcp}.

The other baseline policies, \poln{Interface 2} and \poln{Round Robin}, unsurprisingly 
show a penalty in about 70\% of cases as in most network scenarios Interface~2 has a much higher RTT than Interface~1.

Figure~\ref{fig:speedup_cdf_100_zoom} shows the speedups between 1 and 5 from Figure~\ref{fig:speedup_cdf_100} in more detail. From our data, we find that \poln{EAF} was up
to 2x faster than \poln{Interface 1} in about 23\% of the cases and from 2 to 5x faster
in about 11\% of the cases. We even see speedups of more than 5x in 8.5\% of the cases.
\poln{EAF\_MPTCP} and \poln{MPTCP if1} shows negligibly higher speedups than \poln{EAF}.
Overall, all three policies perform similarly and can take serious advantage of combining multiple access networks.
Even with similar benefits, \poln{EAF} has the advantage over MPTCP that it does not need to be supported by the server and that it cannot be blocked by middleboxes.

Finally, \fref{fig:speedup_cdf_1000} shows the ECDF of the speedups against
\poln{Interface 1} for the Alexa~1000. These look similar to the ones for
Alexa~100 in \ref{fig:speedup_cdf_100}. This gives us confidence that our benefits are
stable for a wide variety of different Web pages.

\subsection{Benefits of Using the Socket Intents Prototype with MPTCP}
\label{sub:eval_mptcp}

\begin{figure}[t!]
  \includegraphics[width=\columnwidth, clip=true, trim=0 20pt 0 54pt]{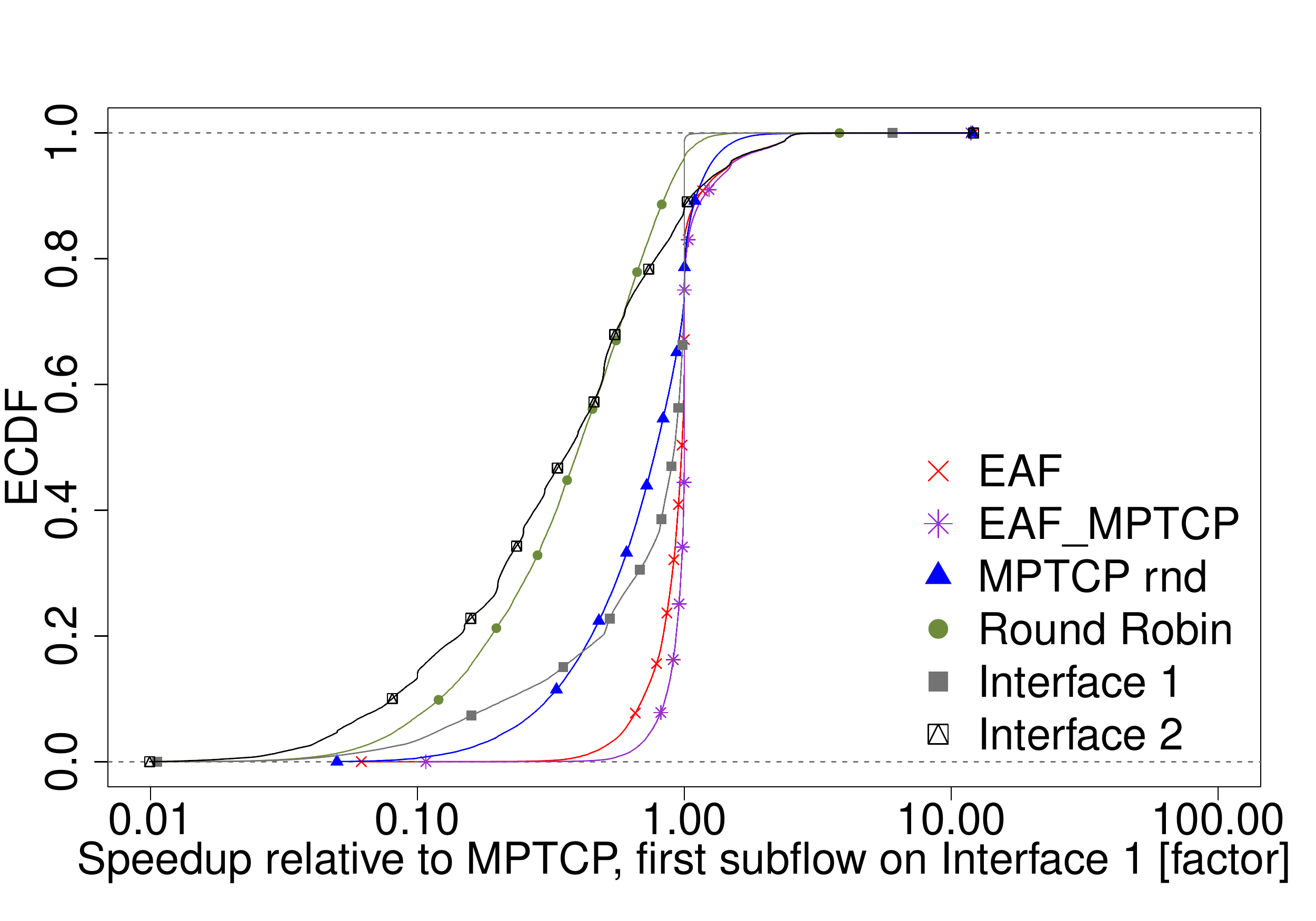}
  \caption{CDF of Speedups vs. \poln{MPTCP if1} for the Alexa Top 100 workload. \label{fig:speedup_cdf_100_mptcp} }
\end{figure}

As described in \sref{sub:eval_muacc}, for our dataset \poln{MPTCP if1} and 
\poln{MPTCP rnd} behave very differently. While both show gains in almost all
cases, \poln{MPTCP rnd} is at a disadvantage in 48\% of the cases while
\poln{MPTCP if1} almost never imposes a penalty.

In \fref{fig:speedup_cdf_100_mptcp}, we compare speedups of our policies for all scenarios and Web pages against \poln{MPTCP if1}. The curves for \poln{EAF} and \poln{EAF\_MPTCP} are close to 1, which means that the page load times are similar to MPTCP in most cases and never considerably worse.
In contrast, if establishing the first subflow for MPTCP over a randomly
chosen interface (\poln{MPTCP rnd}), MPTCP performs worse and can be up to 10x slower
than using \poln{Interface 1} and about 30x slower than \poln{MPTCP if1}.
The reason for that is that Interface 1 has a shorter RTT in most network scenarios. 
As many Web page downloads in our workloads were short and not bandwidth bound,
MPTCP will often perform most of the download over the initial subflow.
Thus, not picking the most suitable one in 50\% of the cases bears a considerable performance penalty.
\poln{EAF\_MPTCP} can always choose the most suitable interface for the first subflow and, therefore, can improve over \poln{MPTCP if1} in cases where Interface~1 is not the  most suitable interface for the first subflow. 
Note that \poln{EAF} shows a similar performance as \poln{MPTCP if1}.
The cases where \poln{EAF} and \poln{EAF\_MPTCP} perform slightly worse than \poln{MPTCP
if1}\footnote{These cases occur because \poln{EAF} and \poln{EAF\_MPTCP} do not take
future transfers into account. They cannot change their decision whether to use MPTCP, while always using MPTCP allows to rebalance traffic between subflows later.}
seem negligible to us given the benefits.

\begin{figure}[t!]
    \subfloat[Interface 1 bandwidth] {
    \includegraphics[width=\columnwidth, clip=true, trim=0 10pt 0 54pt]{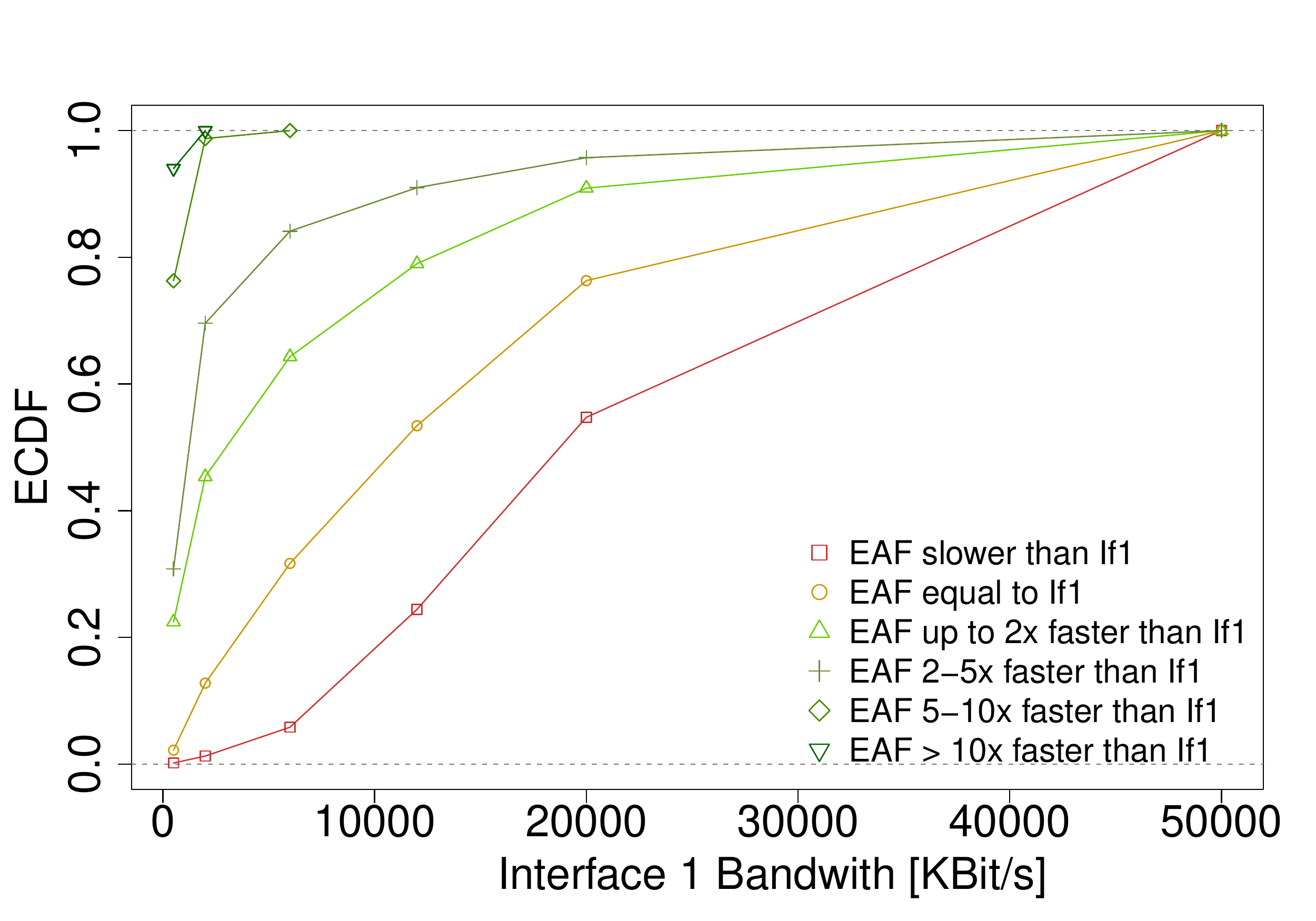}
      \label{fig:if1_bw}
    }
    
    \subfloat[Sizes of Web pages] {
    \includegraphics[width=\columnwidth, clip=true, trim=0 10pt 0 54pt]{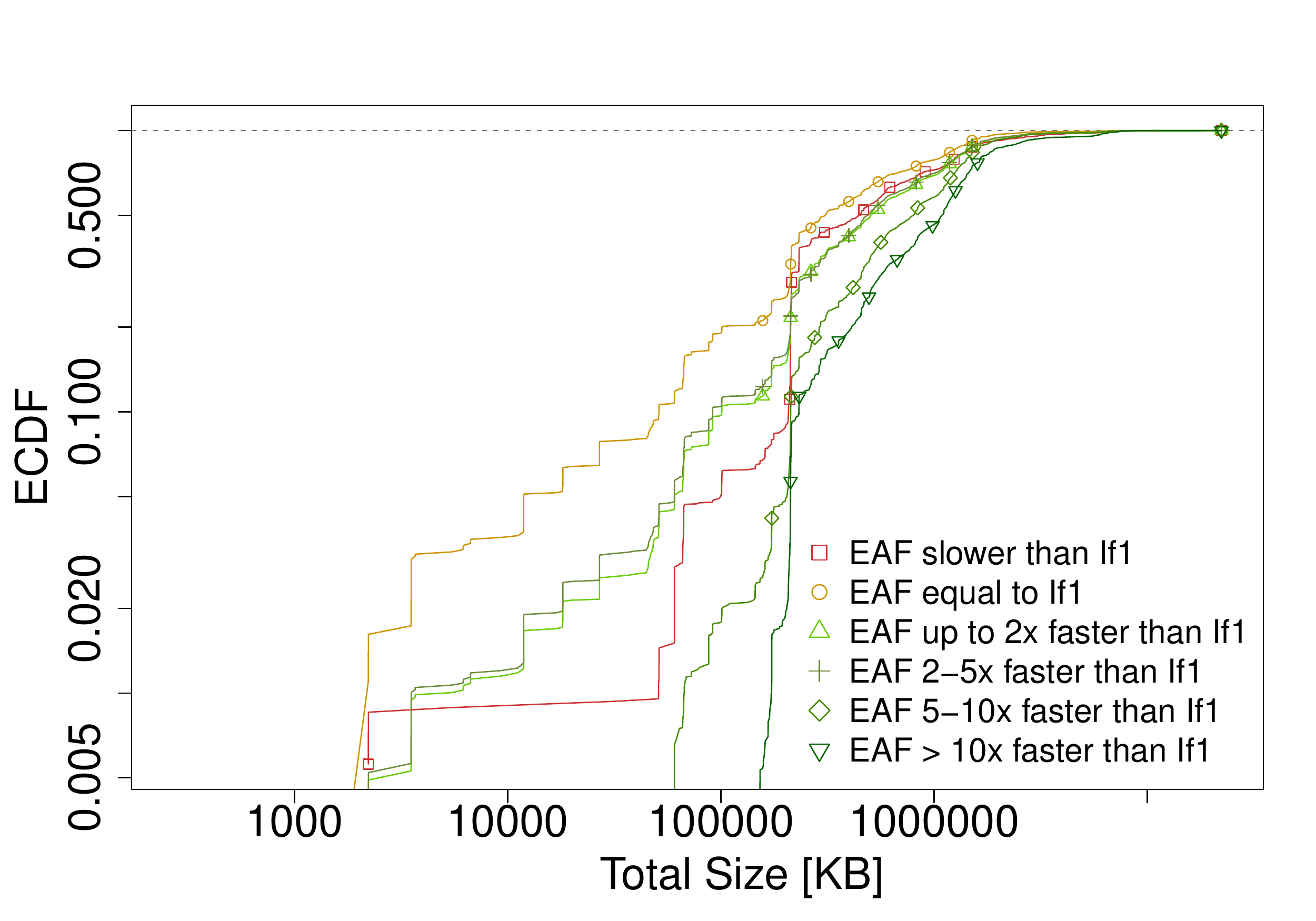}
    \label{fig:total-size}
    }
 \caption{Levels of factors for which we see a certain level of speedup\label{fig:factors} for Alexa Top 100}
\end{figure}

\subsection{Explaining Page Load Time Speedups}
\label{sub:speedup_factors}

To understand how the factors of the scenario and Web page affect the speedups of our policies, we take a closer look at the cases when \poln{EAF} is slower, similar to, or faster than \poln{Interface~1}.

In \fref{fig:factors} we bin the simulation results of \poln{EAF} into six categories
of benefits and show how these distribute among the total Web page sizes and
Interface~1 bandwidths.
Note that these categories contain different numbers of observations, i.e., \emph{EAF is slower} for just 1.5\% of all cases while \emph{EAF is equal to \poln{Interface 1}} for 56.6\% of all cases.

The CDF in Figure~\ref{fig:if1_bw} shows the frequency of the speedup categories over
the different levels of Interface~1 bandwidths from \tref{tab:simulation_levels}.
In cases when \poln{EAF} was slower or equal to \poln{Interface 1}, higher values 
for the Interface~1 bandwidth are more prevalent, while high speedups mostly occur when the Interface~1 bandwidth is low. 
Similarly, we tend to see high speedups for higher levels of Interface~2 bandwidth and for lower levels of Interface~2 RTT (plots omitted).

To explore what kind of Web pages can benefit from our \poln{EAF} policy, we plot the CDF of the speedup categories over the total Web page size in \fref{fig:total-size}. As high speedups occur much more frequently for large Web pages, we conclude that unsurprisingly these take most advantage of using multiple access networks. For the median object size and the number of objects in a Web page we see similar results, with high speedups occurring more frequently in cases with high median object sizes.

Both analyses show that our multi-access policies are most useful when Web page download is bandwidth limited.

\section{Related Work}
\label{sec:related}

We next review related work regarding 
multipath support in general and on the end host's Operating System (OS) in particular.
We then focus on how application needs are taken into account.
Finally, we discuss
the benefits of using multiple access networks in the context of WiFi offloading and MPTCP.

\subsubsection{Multipath} 

For a comprehensive survey of network layer multipath solutions see Qadir et
al.~\cite{qadir2015exploiting}. They present a detailed analysis of the design
choices of how to compute and select routes as well as how to split the flow
across the chosen paths. 

A survey of multipath approaches in some current OSes~\cite{rfc6419}
points out several problems that we also discuss in \sref{sec:bsdsockets_challenges}.
Many OSes support mechanisms for source and destination address selection for
IPv6 multihoming~\cite{rfc5014, rfc6724} and there are proposed Socket API
extensions that enable applications to set preferences \cite{rfc5014}. However,
these address selection algorithms focus on reachability, while we consider
bandwidth aggregation and performance improvement.
Some OSes implement a central connection manager to choose the appropriate
access network, as is also proposed in current research such as by Kiefer et
al.~\cite{kiefer2014feast}. The latter uses policies controlled by the
application and the user and relies on observations of the current network
performance, which is similar to our \polmgr. However, it only works on a
per-flow basis and not per communication unit. Also, their application policies
specify flow prioritizations and constraints, but not different characteristics
of the traffic.

\subsubsection{Application Needs}

Previous work where an application can specify its requirements and
needs often focuses on QoS, e.g., QSockets~\cite{qsockets}. We use the best-effort
approach of \socketintents.  The term \emph{Intents} has its origin in
Intentional Networking~\cite{intentional-networking}, an attempt to explore
mobile network diversity by letting applications specify traffic
characteristics via an extended Socket API. However, they use a per-packet
approach, which introduces complications and overhead, while we use a
per-socket/per-flow or per-request approach. Moreover, they imply guarantees
while we suggest best-effort. Other approaches include ideas from machine
learning to guide application choices, e.g., Deng et
al~\cite{Deng:2014:YNB:2565585.2565588}.

\subsubsection{Socket APIs}

Alternative socket APIs move parts of the application logic to the socket API,
e.g., by requesting a service rather than a protocol, port, and address in the
protocol-independent transport API~\cite{piapi} 
or by exposing all protocols and auxiliary information of the
application in a tree-like structure~\cite{PostSocketsFIT2017}.  
In contrast, our \prototype takes transport protocols as given. Thus, the idea of \socketintents is complementary to the before-mentioned work.

\subsubsection{Offloading}
Multi-access connectivity enables one to balance traffic, e.g., from the mobile
network to the WiFi---offloading---or from WiFi to the mobile
network---onloading. Recently, both variants have gotten a lot of attention in
the research community,
e.g.,~\cite{aijaz2013survey,lee2010mobile,balasubramanian2010augmenting,vallina2012david},
as well as in industry, e.g.,~\cite{lee2014economics,cisco-mdo}.  For a survey
on offloading, we refer to, e.g., Aijaz et al.~\cite{aijaz2013survey}. For a
summary of multi-access connectivity, we refer to, e.g., Schmidt et
al.~\cite{muacc-csws12}.  Examples of recent work on offloading include the
work by Lee et al.~\cite{lee2010mobile}, who demonstrate via a quantitative
study the performance benefit of offloading 3G mobile data to WiFi networks,
and Balasubramanian et al.~\cite{balasubramanian2010augmenting}, who propose
Wiffler to augment mobile 3G capacity with WiFi. For onloading, we, e.g., point
to Vallina et al.~\cite{vallina2012david}. For an analysis of the economics
of offloading see Lee et al.~\cite{lee2014economics}.
Offloading typically implements support for using multiple access networks
within the network or on the application layer, while we provide support for it
within the end-host OS.

\subsubsection{MPTCP}

There have been many studies exploring how an end host can benefit from multiple paths using MPTCP.
Chen et al.~\cite{chen2013measurement} evaluate MPTCP performance in the wild
by comparing its use over a home WiFi network and several different cellular
providers to the use of a single path.  They find that for small files, using a
single path over WiFi is best, while larger files benefit from MPTCP's
aggregated bandwidth.  This observation is shared by Raiciu et
al.~\cite{raiciu2012hard} and Deng et al.~\cite{deng2014wifi}, who emphasize
that the choice of the interface to establish the first subflow is important,
which is in line with our observations.

Han et al.~\cite{han2015anatomy} evaluate page load times of HTTP and SPDY over
WiFi and LTE using a proxy-based setup. They find that SPDY over MPTCP is
always beneficial. This is in contrast to HTTP over MPTCP which in some cases performs
even worse than plain TCP.
Similarly, Nikravesh et al.~\cite{nikravesh2016MPTCP} observe a performance
penalty and energy consumption overhead for apps with small flows in the wild.
As a solution, they propose a proxy with persistent connections over multiple
paths.

A similar approach for controlling MPTCP (see \sref{sub:socket_intents_mptcp})
has been proposed by Hartung and Milind~\cite{hartung2015policy} to 
use MPTCP for LTE bandwidth management.

\subsubsection{SCTP}

Dreibholz et al. propose an advanced stream scheduling policy for SCTP\cite{dreibholz2010transmission} and achieve performance benefits in asymmetric path scenarios using a simulation.
It would be possible to integrate such an advanced scheduler into our policies in future work.

\enlargethispage{0.8mm}

\section{Conclusion}
\label{sec:summary}

Our \prototype is a system to provide seamless OS support for multiple access networks. 
It achieves this goal by allowing applications to express their intents and expectations
for each communication unit, e.g., HTTP request or TCP connection, toward the OS.
The OS then can match the applications' needs and the diverse access networks available 
in a best effort way. 

We evaluate \socketintents in a testbed and confirm that our \prototype can speed up page load time by up to 50\% in typical scenarios.
We then use a simulator to explore speedups across a wide range of Web pages and network characteristics.
Our simulations demonstrate that \socketintents with the \poln{EAF} policy can
improve Web page load time in about 50\% of the cases without incurring
penalties. Therefore, without kernel modification, we can achieve about the same
speedups possible by using MPTCP.
In cases where the access networks characteristics are very different, our
\poln{EAF\_MPTCP} helps MPTCP pick the right interface for the initial subflow and
therefore prevents performance penalties that can occur from using the "`wrong"' 
interface.

\socketintents are a first step for sharing information between applications, the OS,
and the network. This idea is not limited to the use case of multiple access
networks, but can also be beneficial to automatically choose among transport protocols
and can give valuable input to traffic management systems for datacenter networks.
Therefore, we are trying to contribute the key ideas of \socketintents to the IETF 
\emph{TAPS working group}~\cite{I-WG-C.charter-ietf-taps-01} as means to address 
the complexity arising from today's transport layer diversity.

\section*{Acknowledgements}

This work has been supported by Leibniz Prize project funds of DFG - German Research Foundation: Gottfried Wilhelm Leibniz-Preis 2011 (FKZ FE 570/4-1).

Thanks to Thomas Zinner for his feedback and support during the revision of this paper.

\bibliographystyle{IEEEtran/bibtex/IEEEtran}
\bibliography{main,rfc}

\end{document}